\documentclass[11pt]{article}
\usepackage{psfig}
\begin{document}

\fontsize{11}{14.5pt}\selectfont

\begin{center}

{\small Technical Report No.\ 2005,
 Department of Statistics, University of Toronto}

\vspace*{0.65in}

{\huge \bf Slice Sampling} \\[16pt]

{\large Radford M. Neal}\\[2pt]
 Department of Statistics and Department of Computer Science \\
 University of Toronto, Toronto, Ontario, Canada \\
 \texttt{http://www.cs.utoronto.ca/$\sim$radford/} \\
 \texttt{radford@stat.utoronto.ca}\\[10pt]

  29 August 2000
\end{center}

\vspace{8pt} 

\noindent \textbf{Abstract.} Markov chain sampling methods that
automatically adapt to characteristics of the distribution being
sampled can be constructed by exploiting the principle that one can
sample from a distribution by sampling uniformly from the region under
the plot of its density function.  A Markov chain that converges to
this uniform distribution can be constructed by alternating uniform
sampling in the vertical direction with uniform sampling from the
horizontal `slice' defined by the current vertical position, or more
generally, with some update that leaves the uniform distribution over
this slice invariant.  Variations on such `slice sampling' methods are
easily implemented for univariate distributions, and can be used to
sample from a multivariate distribution by updating each variable in
turn.  This approach is often easier to implement than Gibbs sampling,
and more efficient than simple Metropolis updates, due to the ability
of slice sampling to adaptively choose the magnitude of changes made.
It is therefore attractive for routine and automated use.  Slice
sampling methods that update all variables simultaneously are also
possible.  These methods can adaptively choose the magnitudes of
changes made to each variable, based on the local properties of the
density function.  More ambitiously, such methods could potentially
allow the sampling to adapt to dependencies between variables
by constructing local quadratic approximations.  Another approach is
to improve sampling efficiency by suppressing random walks.  This can
be done using `overrelaxed' versions of univariate slice sampling
procedures, or by using `reflective' multivariate slice sampling
methods, which bounce off the edges of the slice.


\section{Introduction}\label{sec-intro}\vspace*{-10pt}

Markov chain methods such as Gibbs sampling (Gelfand and Smith 1990)
and the Metropolis algorithm (Metropolis, {\em et al\/} 1953, Hastings
1970) can be used to sample from many of the complex, multivariate
distributions encountered in statistics.  However, to implement Gibbs
sampling, one may need to devise methods for sampling from
non-standard univariate distributions, and to use the Metropolis
algorithm, one must find an appropriate `proposal' distribution that
will lead to efficient sampling.  The need for such special tailoring
limits the routine use of these methods, and inhibits the development
of software that automatically constructs Markov chain samplers from
model specifications.  Furthermore, many common Markov chain samplers
are inefficient, due to a combination of two flaws.  First, they may
try to make changes that are not well adapted to the local properties
of the density function, with the result that changes must be made in
small steps.  Second, these small steps take the form of a random
walk, in which about $n^2$ such steps are needed in order to move a
distance that could be traversed in only $n$ steps if these steps
moved consistently in one direction.

In this paper, I describe a class of `slice sampling' methods that can
be applied to a wide variety of distributions.  Simple forms of
univariate slice sampling are an alternative to Gibbs sampling that
avoids the need to sample from non-standard distributions.  These
slice sampling methods can adaptively change the scale of changes
made, which makes them easier to tune than Metropolis methods, and
also avoids problems that arise when the appropriate scale of changes
varys over the distribution.  More complex slice sampling methods can
adapt to the dependencies between variables, allowing larger changes
than would be possible with Gibbs sampling or simple Metropolis
methods.  Slice sampling methods that improve sampling by suppressing
random walks can also be constructed.  

Slice sampling originates with the observation that one can sample
from a univariate distribution by sampling points uniformly from the
region under the curve of its density function, and then looking only
at the horizontal coordinates of the sample points.  A Markov chain
that converges to this uniform distribution can be constructed by
alternately sampling uniformly from the vertical interval defined by
the density at the current point, and from the union of intervals that
constitutes the horizontal `slice' though the plot of the density
function that this vertical position defines.  If this last step is
still difficult, one may substitute some other update that leaves the
uniform distribution over the current slice invariant.  To sample from
a multivariate distribution, such single-variable slice sampling
updates can be applied to each variable in turn.  The details of these
single-variable slice sampling methods are described in
Section~\ref{sec-sing}.

One can also apply the slice sampling approach to a multivariate
distribution directly, as described in Section~\ref{sec-multi}, by
sampling uniformly under the multidimensional plot of its density
function.  As for a univariate distribution, this can be done by
alternately sampling uniformly from the vertical interval from zero up
to the density at the current point, and then uniformly from the slice
defined by this vertical position.  When the slice is
high-dimensional, how to efficiently sample from it is less obvious
than for single-variable slice sampling, but one gains the possibility
of sampling in a way that respects the dependencies between variables.
I show how, in the context of slice sampling, the way changes are
proposed can be adapted to respect these dependencies, based on local
information about the density function.  In particular, local
quadratic approximations could be constructed, as have been used very
successfully for optimization problems.  Adaptive slice sampling
appears to be simpler than a somewhat analogous scheme proposed for
the Metropolis algorithm (Mira 1998, Chapter~5; Tierney and Mira 1999;
Green and Mira 1999).  However, further research will be needed to
fully exploit the adaptive capabilities of multivariate slice
sampling.

One might instead accept that dependencies between variables will lead
to the distribution being explored in small steps, but try at least to
avoid exploring the distribution by an inefficient random walk, which
is what happens when simple forms of the Metropolis algorithm are
used.  The benefits of random walk suppression are analysed
theoretically in some simple contexts by Diaconis, Holmes, and Neal
(in press).  Large gains in sampling efficiency can be obtained in
practice when random walks are suppressed using the Hybrid Monte Carlo
or other dynamical methods (Duane, Kennedy, Pendleton, and Roweth
1987; Horowitz 1991; Neal 1994; and for reviews from a more
statistical perspective, Neal 1993, 1996), or by using an
overrelaxation method (Adler 1981; Barone and Frigessi 1990; Green and
Han 1992; Neal 1998).  Dynamical and overrelaxation methods are not
always easy to apply, however.  Use of Markov chain samplers that
avoid random walks would be assisted by the development of methods
that require less special programming and parameter tuning.

Two approaches to random walk suppression based on slice sampling are
discussed in this paper.  In Section~\ref{sec-over}, I show how one
can implement an overrelaxed version of the single-variable slice
sampling scheme.  This may provide the benefits of Adler's (1981)
Gaussian overrelaxation method for more general distributions.  In
Section~\ref{sec-reflect}, I describe slice sampling analogues of
dynamical methods, which move around a multi-variable slice using a
stepping procedure that proceeds consistently in one direction while
reflecting off the slice boundaries.  Although these more elaborate
slice sampling methods require more tuning than the single-variable
slice sampling schemes, they may still be easier to apply than
alternative methods that avoid random walks.

To illustrate the advantages of the adaptive nature of slice sampling,
I show in Section~\ref{sec-demo} how it can help avoid disaster when
sampling from a distribution that is typical of priors for
hierarchical Bayesian models.  Simple Metropolis methods can give the
wrong answer for this problem, while providing little indication that
anything is amiss.

This paper concludes (in Section~\ref{sec-disc}) with a discussion of
the merits of the various slice sampling methods in comparison with
other Markov chain methods, and of their suitability for routine and
automated use.  Below, I set the stage by discussing general-purpose
Markov chain methods that are currently in wide use.  Readers who are
quite familiar with Markov chain sampling and are eager to get to the
main idea can skip immediately to Section~\ref{sec-idea}.

\section{General-purpose Markov chain sampling 
         methods}\label{sec-gp}\vspace*{-10pt}

Applications of Markov chain sampling in statistics often involve
sampling from many distributions.  In Bayesian applications, we must
sample from the posterior distribution for the parameters of a model
given certain data.  Different datasets will produce different
posterior distributions, which may differ in important characteristics
such as diffuseness and multimodality.  Furthermore, we will often
wish to consider a variety of models.  For routine use of Markov chain
methods, it is important to minimize the amount of effort that the
data analyst must spend in order to sample from all these
distributions.  Ideally, a Markov chain sampler would be constructed
automatically for each model and dataset.

The Markov chain method most commonly used in statistics is Gibbs
sampling, popularized by Gelfand and Smith (1990).  Suppose that
we wish to sample from a distribution over $n$ state variables (eg,
model parameters), written as $x = (x_1,\ldots,x_n)$, with probability
density $p(x)$.  Gibbs sampling proceeds by sampling in succession
from the conditional distributions for each $x_i$ given the current
values of the other $x_j$ for $j \ne i$, with conditional densities
written as $p(x_i\,|\,\{x_j\}_{j \ne i})$.  Repetition of this
procedure defines a Markov chain which leaves the desired distribution
invariant, and which in many circumstances is ergodic (eg, when
$p(x)>0$ for all $x$).  Running the Gibbs sampler for a sufficiently
long time will then produce a sample of values for $x$ from close to
the desired distribution, from which the expectations of quantities of
interest (eg, posterior means of parameters) can be estimated.

Gibbs sampling can be done only if we know how to sample from all the
required conditional distributions.  These sometimes have standard
forms for which efficient sampling methods have been developed, but
there are many models for which sampling from these conditional
distributions requires the development of custom algorithms, or is
infeasible in practice (eg, for multilayer perceptron networks (Neal
1996)).  Note, however, that once methods for sampling from these
conditional distributions have been found, no further tuning
parameters need be set in order to produce the final Markov chain
sampler.

The routine use of Gibbs sampling has been assisted by the development
of Adaptive Rejection Sampling (ARS) (Gilks and Wild 1992; Gilks
1992), which can be used to efficiently sample from any conditional
distribution whose density function is log concave, given only the
ability to compute some function, $f_i(x_i)$, that is proportional to
the conditional density, $p(x_i\,|\,\{x_j\}_{j \ne i})$ (the ability
to also compute the derivative, $f_i^{\prime}(x_i)$, is helpful, but
not essential).  This method has been used for some time by the BUGS
software (Thomas, Spiegelhalter, and Gilks 1992) to automatically
generate Markov chain samplers from model specifications.  The first
step in applying ARS is to find points on each side of the mode of the
conditional distribution (one of which can be the current point).
This will in general require a search, which will in turn require the
choice of some length scale for an initial step.  However, the burden
of setting this scale parameter is lessened by the fact that a good
value for it can be chosen `retrospectively', based on past iterations
of the Markov chain, without invalidating the results (since the
setting of this parameter affects only the computation time, not the
distribution sampled from).

The Adaptive Rejection Metropolis Sampling (ARMS) method (Gilks, Best,
and Tan 1995) generalizes ARS to conditional distributions whose
density functions may not be log-concave.  However, when the density
is not log-concave, ARMS does not produce a new point drawn
independently from the conditional distribution, but merely updates
the current point in a fashion that leaves this distribution
invariant.  Applying ARMS to sample from the conditional distribution
of each variable in succession will result in an equilibrium
distribution that is exactly correct, but when some conditional
distributions are not log-concave, it may take longer to approach this
equilibrium than would be the case if true Gibbs sampling were used.
Also, when a conditional distribution is not log-concave, the points
used to set up the initial approximation to it must not be chosen with
reference to past iterations, as this could result in the wrong
distribution being sampled (Gilks, Neal, Best, and Tan 1997).  The
initial approximation must be chosen based only on prior knowledge
(including any preliminary Markov chain sampling runs), and on the
current values of the other variables.  Unlike ARS, neither the
current value of the variable being updated, nor any statistics
collected from previous updates (eg, the typical scale of changes) can
be used.  This hinders routine use of the method.

Another general way of constructing a Markov chain sampler is to
perform Metropolis updates (Metropolis, {\em et al\/} 1953, Hastings
1970), either for each variable in turn, as with Gibbs sampling, or
for all variables simultaneously.  A Metropolis update starts with the
random selection of a `candidate' state, drawn from a `proposal'
distribution.  The candidate state is then accepted or rejected as the
new state of the Markov chain, based on the ratio of the probability
densities of the candidate state and the current state.  If the
candidate state is rejected, the new state is the same as the old
state.

A simple `random-walk' Metropolis scheme can be constructed based on a
symmetric proposal distribution (eg, Gaussian) that is centred on the
current state.  All variables could be updated simultaneously in such
a scheme, or alternatively, one variable could be updated at a time.
In either case, a scale parameter is required for each variable to fix
the width of the proposal distribution in that dimension.  For the
method to be valid, these scale parameters must not be set on the
basis of past iterations, but rather only on the basis of prior
knowledge (including preliminary runs), and the current values of
variables that are not being changed in the present update.  Choosing
too large a value for the scale of a proposal distribution will result
in a high rejection rate, while choosing too small a value will result
in inefficient exploration via a random walk with unnecessarily small
steps.  Furthermore, the appropriate scale for Metropolis proposals
may vary from one part of the distribution to another, in which case
no single value will produce acceptable results.  Selecting a scale at
random from some range can sometimes alleviate these problems, but at
a large cost in wasted effort when the scale selected is inappropriate.

It is tempting to tune the Metropolis proposal distribution based on
the rejection rate in past iterations of the Markov chain, but such
`retrospective tuning' is not valid in general, since it can disturb
the stationary distribution to which the process converges (as was
also the case for ARMS).  Fixing the proposal distribution based on a
preliminary run is allowed, but if the original proposal distribution
was not good, such a preliminary run may not have sampled from the
whole distribution, and hence may be a bad guide for tuning.

We therefore see that although Gibbs sampling and Metropolis methods
have been used to do much useful work, there is a need for better
methods, that can be routinely applied in a wider variety of
situations.  One of my objectives in this paper is to find variations
on slice sampling that can be used to sample from any continuous
distribution, given only the ability to evaluate a `black-box'
function that is proportional to its density, and in some cases, to
also evaluate the gradient of the log of this function.  For many
distributions, these new methods will not sample more efficiently than
true Gibbs sampling or a well-designed Metropolis scheme, but the
slice sampling methods will often requiring less effort to implement
and tune.  For some distributions, however, slice sampling can be much
more efficient, because it can adaptively choose a scale for changes
appropriate for the region of the distribution currently being
sampled.  Slice samplers that adapt in more elaborate ways, or that
are designed to suppress random walks, can potentially be much faster
than simple Metropolis methods or Gibbs sampling.

\section{The idea of slice sampling}\label{sec-idea}\vspace*{-10pt}

Suppose we wish to sample from a distribution for a variable, $x$,
taking values in some subset of
$\Re^n$, with density function proportional to some function $f(x)$.
We can do this by sampling uniformly from the \mbox{$n\!+\!1$}
dimensional region that lies under the plot of $f(x)$.  This idea can
be formalized by introducing an auxiliary real variable, $y$, and
defining a joint distribution over $x$ and $y$ that is uniform over
the region \mbox{$U = \{\,(x,y)\ :\ 0 < y < f(x)\,\}$} below
the curve or surface defined by $f(x)$.  That is, the joint density 
for $(x,y)$ is
\begin{eqnarray}
  p(x,y) & = & 
     \left\{\begin{array}{ll}
         1/Z & \mbox{~~if $0<y<f(x)$} \\ 0 & \mbox{~~otherwise}
     \end{array}\right.\label{eq-joint}
\end{eqnarray}
where $Z=\int f(x)\,dx$. The marginal density for $x$ is then
\begin{eqnarray}
  p(x) & = & \int_0^{f(x)} (1/Z)\, dy \ \ = \ \ f(x)\,/\,Z
\end{eqnarray}
as desired.  To sample for $x$, we can sample jointly for $(x,y)$, and
then ignore $y$.

Generating independent points drawn uniformly from $U$ may not be
easy, so we might instead define a Markov chain that will converge to
this uniform distribution.  Gibbs sampling is one possibility: We
sample alternately from the conditional distribution for $y$ given the
current $x$ --- which is uniform over the interval $(0,f(x))$ --- and
from the conditional distribution for $x$ given the current $y$ ---
which is uniform over the region \mbox{$S = \{\, x\,:\,y < f(x)\,\}$},
which I call the `slice' defined by $y$.  Generating an independent
point drawn uniformly from $S$ may still be difficult, in which case
we can substitute some update for $x$ that leaves the uniform
distribution over $S$ invariant.

Similar auxiliary variable methods been used in the past.  Higdon
(1996) has interpreted the standard Metropolis algorithm in these
terms.  The highly successful Markov chain algorithm for the Ising
model due to Swendsen and Wang (1987) can also be seen as an auxiliary
variable method, which has been generalized by Edwards and Sokal
(1988).  In their scheme, the density (or probability mass) function
is proportional to a product of $k$ functions: \mbox{$p(x) \propto
f_1(x) \cdots f_k(x)$}.  They introduce $k$ auxiliary variables,
$y_1,\ldots,y_k$, and define a joint distribution for
$(x,y_1,\ldots,y_k)$ which is uniform over the region where $0 < y_i <
f_i(x)$ for $i=1,\ldots,k$.  Gibbs sampling, or some related Markov
chain procedure, can then be used to sample for $(x,y_1,\ldots,y_k)$,
much as described above for the case of a single auxiliary variable.
Applications of such methods to image analysis have been discussed by
Besag and Green (1993) and by Higdon (1996).

Mira and Tierney (in press) have shown that these auxiliary variable
methods, with one or with many auxiliary variables, are uniformly
ergodic under certain conditions.  Roberts and Rosenthal (1999) have
shown that these methods are geometrically ergodic under weaker
conditions, and have also found some quantitative convergence bounds.
These results all assume that the sampler generates a new value for
$x$ that is uniformly drawn from $S$, independently of the old value,
which is often difficult in practice.

Concurrently with the work reported here\footnote{
  An earlier version of this paper, under the title ``Markov chain Monte Carlo 
  methods based on `slicing' the density function'' was issued in November 1997
  as Technical Report 9722, Department of Statistics, University of Toronto
  (available from my web page).  It contains essentially all the material in
  this paper with the exception of Sections 5 and 8.
  The first application of the methods developed here, by Frey (1997), predates
  this technical report, referencing it as being `in preparation'.},
Damien, Wakefield, and Walker (in press) have viewed methods based on
multiple auxiliary variables as a general approach to constructing
Markov chain samplers for Bayesian inference problems.  They
illustrate how one can often decompose $f(x)$ into a product of $k$
factors for which the intersection of the sets
$\{\,x\,:\,y_i < f_i(x)\,\}$ is easy to compute.  This leads
to an easily implemented sampler, but convergence is slowed by the
presence of many auxiliary variables.  For example, for a model of $k$
i.i.d.\ data points, one simple approach (similar to some examples of
Damien, \textit{et al}) is to have a factor (and auxiliary variable)
for each data point, with the product of these factors being the
likelihood. (Suppose for simplicity that the prior is uniform, and so
needn't be represented in the posterior density.)  For many models,
finding $\{\,x\,:\,y_i < f_i(x)\,\}$ will be easy to compute when
$f_i$ is the likelihood from one data point.  However, if this
approach is applied to $n$ data points that are modeled as coming
from a Gaussian distribution with mean $\mu$ and variance 1, it is
easy to see that after the $y_i$ are chosen, the allowable range for
$\mu$ will have width of order $1/n$.  Since the width of the posterior
distribution for $\mu$ will be of order $1/\sqrt{n}$, and since the posterior
will be explored by a random walk, the convergence time will be of order
$n$.  Gibbs sampling would, of course, converge in a single
iteration when there is only one parameter, and the slice sampling
methods of this paper would also converge very rapidly for this
problem, for any $n$.  Using a large number of auxiliary variables is a 
costly way to avoid difficult computations.

I therefore am concerned in this paper with methods based on slice
sampling with a single auxiliary variable.  So that these methods will
be practical for a wide range of problems, they often use updates for
$x$ that do not produce a point drawn independently from the slice,
$S$, but merely change $x$ in some fashion that leaves the uniform
distribution over $S$ invariant.  This allows the methods to be used
for any continuous distribution, provided only that we can compute
some function, $f(x)$, that is proportional to the density.

\section{Single-variable slice sampling methods}\label{sec-sing}\vspace*{-10pt}

Slice sampling is simplest when only one (real-valued) variable is
being updated.  This will of course be the case when the distribution
of interest is univariate, but more typically, the single-variable
slice sampling methods of this section will be used to sample from
a multivariate distribution for $x = (x_1,\ldots,x_n)$ by sampling
repeatedly for each variable in turn.  To update $x_i$, we must be
able to compute a function, $f_i(x_i)$, that is proportional to
$p(x_i\,|\,\{x_j\}_{j \ne i})$, where $\{x_j\}_{j \ne i}$ are the
values of the other variables.  Often, the joint distribution for
$x_1,\ldots,x_n$ will be defined by some function,
$f(x_1,\ldots,x_n)$, that is proportional to the joint density, in
which case we can simply take $f_i(x_i) = f(\ldots,x_i,\ldots)$, where
the variables other than $x_i$ are fixed to their current values.

To simplify notation, I will here write the single real variable being
updated as $x$ (with subscripts denoting different such points, not
components of $x$).  I will write $f(x)$ for the function proportional to
the probability density of $x$.  The single-variable slice sampling methods
discussed here replace the current value, $x_0$, with a new value, $x_1$,
found by a three-step procedure:
\begin{enumerate}\vspace*{-8pt}
\item[\textit{a)}] 
          Draw a real value, $y$, uniformly from $(0,f(x_0))$, thereby defining
          a horizontal `slice': \mbox{$S = \{\, x\,:\,y < f(x)\,\}$}.  Note
          that $x_0$ is always within $S$.
\item[\textit{b)}] 
          Find an interval, $I = (L,R)$, around $x_0$ that
          contains at least a big part of the slice.\vspace*{-2pt}
\item[\textit{c)}] 
          Draw the new point, $x_1$, from the part of the
          slice within this interval (ie, from $S \cap I$).\vspace*{-8pt}
\end{enumerate}

Step \textit{(a)} picks a value for the auxiliary variable that is
characteristic of slice sampling.  Note that there is no need to
retain this auxiliary variable from one iteration of the Markov chain
to the next, since its old value is forgotten at this point anyway.
In practice, it is often safer to compute $g(x) = \log(f(x))$ rather
than $f(x)$ itself, in order to avoid possible problems with
floating-point underflow.  One can then use the auxiliary variable
\mbox{$z = \log(y) = g(x_0)-e$}, where $e$ is exponentially distributed
with mean one, and define the slice by \mbox{$S = \{\, x\,:\, z <
g(x)\, \}$}.  

\begin{figure}[p]

\vspace*{-10pt}

\centerline{\psfig{figure=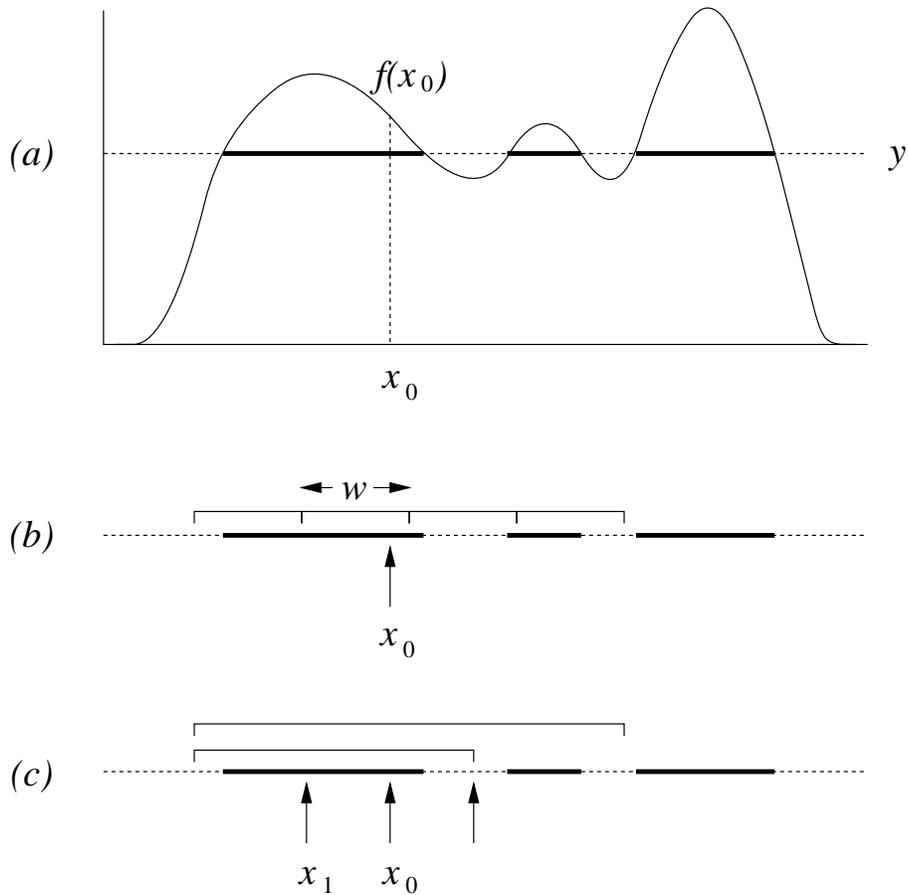}}

\caption[]{A single-variable slice sampling update using the
stepping-out and shrinkage procedures.  A new point, $x_1$, is
selected to follow the current point, $x_0$, in three steps.
\textit{(a)} A vertical level, $y$, is drawn uniformly from
$(0,f(x_0))$, and used to define a horizontal `slice', indicated in
bold.  \textit{(b)} An interval of width $w$ is randomly
positioned around $x_0$, and then expanded in steps of size $w$ until 
both ends are outside the slice.  \textit{(c)} A new point, $x_1$, 
is found by picking uniformly from the interval until a point inside 
the slice is found.  Points picked that are outside the slice are 
used to shrink the interval.}\label{fig-sing}

\end{figure}

\begin{figure}[p]

\vspace*{10pt}

\centerline{\psfig{figure=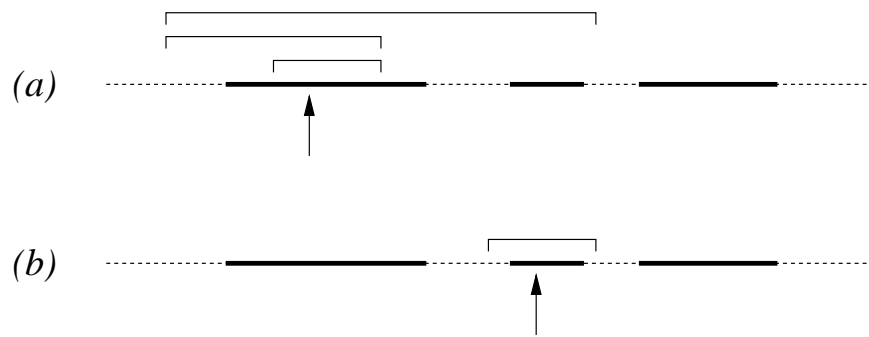}}

\caption[]{The doubling procedure.  In \textit{(a)}, the initial
interval is doubled twice, until both ends are outside the slice.
In \textit{(b)}, where the start state is different, no doubling is done.
}\label{fig-dblp}

\end{figure}

Steps \textit{(b)} and \textit{(c)} can potentially be implemented in
several ways, which must of course be such that the resulting Markov
chain leaves the distribution defined by $f(x)$
invariant. Figure~\ref{fig-sing} illustrates one generally-applicable
method, in which the interval is found by `stepping out', and the new
point is drawn with a `shrinkage' procedure.  Figure~\ref{fig-dblp}
illustrates an alternative `doubling' procedure for finding the
interval.  These and some other variations are described in detail
next, followed by a proof that the resulting transitions leave the
correct distribution invariant.  I then describe some shortcuts
that are possible when the distribution is unimodal.

\subsection{Finding an appropriate interval}\label{sub-int}\vspace*{-8pt}

After a value for the auxiliary variable has been drawn, defining the
slice $S$, the next task is to find an interval $I=(L,R)$, containing
the current point, $x_0$, from which the new point, $x_1$, will be
drawn.  We would like this interval to contain as much of the slice as
is feasible, so as to allow the new point to differ as much as
possible from the old point, but we would also like to avoid intervals
that are much larger than the slice, as this will make the subsequent
sampling step less efficient.

Several schemes for finding an interval are possible:
\begin{enumerate}\vspace*{-8pt}
\item[1)] Ideally, we would set $L = \inf(S)$ and $R = \sup(S)$.  That is,
          we would set $I$ to the smallest interval that contains the whole
          of $S$.  This may not be feasible, however.\vspace*{-2pt}
\item[2)] If the range of $x$ is bounded, we might simply let $I$ be that 
          range.  However, this may not be good if the slice is typically much
          smaller than the range.\vspace*{-2pt}
\item[3)] Given an estimate, $w$, for the scale of $S$, we can randomly
          pick an initial interval of size $w$, containing $x_0$, and
          then perhaps expand it by a `stepping out' procedure.\vspace*{-2pt}
\item[4)] Similarly, we can randomly pick an initial interval of size $w$, and 
          then expand it by a `doubling' procedure.\vspace*{-8pt}
\end{enumerate}

For each scheme, we must also be able to find the set $A$ of acceptable
successor states, defined as follows:
\begin{eqnarray}
  A & = & \{\, x \,:\, x \in S \cap I \mbox{~and~} 
    P(\mbox{Select $I$}\ |\ \mbox{At state $x$}) 
  = P(\mbox{Select $I$}\ |\ \mbox{At state $x_0$})\, \}\label{eq-A}
\end{eqnarray}
That is, $A$ is the set of states from which we would be as likely to
choose the interval $I$ as we were to choose this $I$ from the current
state.  When we subsequently sample from within $I$ (see
section~\ref{sub-samp}), we will ensure that the state chosen is in
$A$, a fact which will be used in the proof of correctness in
section~\ref{sub-correct}.  Clearly, for schemes~(1) and~(2), $A = S$.
For scheme~(3), we will arrange that $A = S \cap I$.  Things are not
so simple for scheme~(4), for which a special test of whether a state
is in $A$ may be necessary.

Scheme (1), in which $I$ is set to the smallest interval containing
$S$, will be feasible when all solutions of $f(x)=y$ can be found
analytically, or by an efficient and robust numerical method, but one
cannot expect this in general.  Often, even the number of disjoint
intervals making up $S$ will be hard to determine.

Scheme (2) is certainly easy to implement when the range of $x$ is
bounded, and one can of course always arrange this by applying a
suitable transformation.  However, if the slice is usually much
smaller than the full range, the subsequent sampling (see
section~\ref{sub-samp}) will be inefficient.  This scheme has been
used by Frey (1997).

\begin{figure}[p]

\begin{tabbing}
  \hspace{60pt} \=\+ \hspace{45pt} \= \hspace{13pt} \= \hspace{10pt} \= \kill
  Input:  \> $f$   \> $=$ \> function proportional to the density \\
          \> $x_0$ \> $=$ \> the current point \\
          \> $y$   \> $=$ \> the vertical level defining the slice \\
          \> $w$   \> $=$ \> estimate of the typical size of a slice \\
          \> $m$   \> $=$ \> integer limiting the size of a slice to $mw$
          \\[8pt]
  Output: \> $(L,R)\ =\ \mbox{the interval found}$
\end{tabbing}\vspace{8pt}

\begin{tabbing}
  \hspace{60pt} \=\+ \hspace{15pt} \= \hspace{14pt} \= \kill
  $U$\> $\sim$\> $\mbox{Uniform}\,(0,1)$ \\
  $L$\> $\leftarrow$\> $x_0\ -\ w*U$ \\
  $R$\> $\leftarrow$\> $L\ +\ w$ \\[8pt]
  $V$\> $\sim$\> $\mbox{Uniform}\,(0,1)$ \\
  $J$\> $\leftarrow$\> $\mbox{Floor}\,(m*V)$ \\
  $K$\> $\leftarrow$\> $(m-1)\ -\ J$
\end{tabbing}\vspace{8pt}
\begin{tabbing}
\hspace{60pt} \= repeat while $J>0$ and $y<f(L)$:
\end{tabbing}\vspace{3pt}
\begin{tabbing}
  \hspace{80pt} \=\+ \hspace{15pt} \= \hspace{14pt} \= \kill
  $L$\> $\leftarrow$\> $L\ -\ w$ \\
  $J$\> $\leftarrow$\> $J-1$
\end{tabbing}\vspace{8pt}
\begin{tabbing}
  \hspace{60pt} \= repeat while $K>0$ and $y<f(R)$:
\end{tabbing}\vspace{3pt}
\begin{tabbing}
  \hspace{80pt} \=\+ \hspace{15pt} \= \hspace{14pt} \= \kill
  $R$\> $\leftarrow$\> $R\ +\ w$ \\
  $K$\> $\leftarrow$\> $K-1$
\end{tabbing}

\caption[]{The `stepping out' procedure for finding an interval around
the current point.  The notation $U\ \sim\ \mbox{Uniform}\,(0,1)$ indicates
that $U$ is set to a number randomly drawn from the uniform distribution on
$(0,1)$.}\label{fig-step}

\end{figure}

\begin{figure}[p]

\begin{tabbing}
  \hspace{60pt} \=\+ \hspace{45pt} \= \hspace{13pt} \= \hspace{10pt} \= \kill
  Input:  \> $f$   \> $=$ \> function proportional to the density \\
          \> $x_0$ \> $=$ \> the current point \\
          \> $y$   \> $=$ \> the vertical level defining the slice \\
          \> $w$   \> $=$ \> estimate of the typical size of a slice \\
          \> $p$   \> $=$ \> integer limiting the size of a slice to $2^pw$
          \\[8pt]
  Output: \> $(L,R)\ =\ \mbox{the interval found}$
\end{tabbing}\vspace{8pt}

\begin{tabbing}
  \hspace{60pt} \=\+ \hspace{15pt} \= \hspace{14pt} \= \kill
  $U$\> $\sim$\> $\mbox{Uniform}\,(0,1)$ \\
  $L$\> $\leftarrow$\> $x_0\ -\ w*U$ \\
  $R$\> $\leftarrow$\> $L\ +\ w$ \\
  $K$\> $\leftarrow$\> $p$ 
\end{tabbing}\vspace{8pt}
\begin{tabbing}
\hspace{60pt} \= repeat while $K>0$ and \{ $y<f(L)$ or $y<f(R)$ \}:
\end{tabbing}\vspace{3pt}
\begin{tabbing}
  \hspace{80pt} \=\+ \hspace{15pt} \= \hspace{14pt} \= \kill
  $V$\> $\sim$\> $\mbox{Uniform}\,(0,1)$ \\[1pt]
  if $V<1/2$ \= then \= $L\ \leftarrow\ L\ -\ (R-L)$ \\
             \> else \> $R\ \leftarrow\ R\ +\ (R-L)$ \\
  \hspace{15pt} \= \hspace{14pt} \= \kill
  $K$\> $\leftarrow$\> $K-1$
\end{tabbing}

\caption[]{The `doubling' procedure for finding an interval around
the current point.  Note that it is possible to save some computation in
second and later iterations of the loop, since only 
one of $f(L)$ and $f(R)$ will have changed from the previous iteration.
}\label{fig-dbl}

\end{figure}

The `stepping out' procedure (scheme (3) above) is appropriate for any
distribution, provided that some rough estimate, $w$, for the typical
width of the slice is available.  The manner in which an interval is
found by stepping out is illustrated in Figure~\ref{fig-sing}(b) and
the procedure is given in detail in Figure~\ref{fig-step}.  The size
of the interval found can be limited to $mw$, for some specified
integer $m$, or the interval can be allowed to grow to any size (ie,
$m$ can be set to infinity), in which case the procedure can be
simplified in an obvious way (eliminating all references to $J$ and
$K$).  Note that the random positioning of the initial interval and
the random apportioning of the maximum number of steps $m$ into a
limit on going to the left and a limit on going to the right are
essential for correctness, as they ensure that the final interval
could equally well have been produced from any point within $S \cap
I$.

If $m$ is set to one in the stepping out procedure, the interval will
always be of size $w$, and there will be no need to evaluate $f$ at
its endpoints.  This saves some computation time, but is undesirable
if $w$ might be much too small.

The `doubling' procedure (scheme (4) above) can expand the interval
faster than the stepping out procedure, and hence may be more
efficient when the estimated size of the slice ($w$) turns out to be
too small.  This procedure is illustrated in Figure~\ref{fig-dblp},
and given in detail in Figure~\ref{fig-dbl}.  Doubling produces a
sequence of intervals, each twice the size of the previous one, until
an interval is found with both ends outside the slice, or a
predetermined limit is reached.  Note that when the interval is
doubled the two sides are not expanded equally.  Instead just one side
is expanded, chosen at random (irrespective of whether that side is
already outside the slice).  This is essential to the correctness of
the method, since it produces a final interval that could have been
obtained from points other than the current one.  The set $A$ of
acceptable next states is restricted to those for which the same
interval could have been produced, and is in general not all of $S
\cap I$.  This complicates the subsequent sampling somewhat, as
described below.

\subsection{Sampling from the part of the slice within the 
            interval}\label{sub-samp}\vspace*{-8pt}

Once an interval, $I=(L,R)$, has been found containing the current
point, $x_0$, the final step of the single-variable slice sampling
procedure is to randomly draw a new point, $x_1$, from within this
interval.  This point must lie within the set $A$ of points acceptable
as the next state of the Markov chain, defined in
equation~(\ref{eq-A}).

Two methods could be used to sample from $I$:
\begin{enumerate}\vspace*{-8pt}
\item[\textit{i)}] Repeatedly sample uniformly from $I$ until a point
                   is drawn that lies within $A$.\vspace*{-2pt}
\item[\textit{ii)}] Repeatedly sample uniformly from an interval that is 
                    initially equal to $I$, and which shrinks each time a 
                    point is drawn that is not in $A$, until a point within
                    $A$ is found.\vspace{-8pt}
\end{enumerate}
Method~\textit{(i)} could potentially be very inefficient, if ever $A$
turns out to be a tiny portion of $I$.  The shrinkage of the interval in
method~\textit{(ii)} ensures that the expected number of points drawn
will not be too large, making this a more appropriate method for general
use.

\begin{figure}[p]

\begin{tabbing}
  \hspace{60pt} \=\+ \hspace{45pt} \= \hspace{13pt} \= \hspace{10pt} \= \kill
  Input:  \> $f$   \> $=$ \> function proportional to the density \\
          \> $x_0$ \> $=$ \> the current point \\
          \> $y$   \> $=$ \> the vertical level defining the slice \\
          \> $w$   \> $=$ \> estimate of the typical size of a slice \\
          \> $(L,R)\ =\ \mbox{the interval to sample from}$ \\[8pt]
  Output: \> $x_1$ \> $=$ \> the new point
\end{tabbing}\vspace{8pt}

\begin{tabbing}
  \hspace{60pt} \=\+ \hspace{15pt} \= \hspace{14pt} \= \kill
  $\bar L\ \leftarrow\ L$,\hspace{6pt} $\bar R\ \leftarrow\ R$
\end{tabbing}\vspace{8pt}
\begin{tabbing}
\hspace{60pt} \= repeat:
\end{tabbing}\vspace{6pt}
\begin{tabbing}
  \hspace{80pt} \=\+ \hspace{15pt} \= \hspace{14pt} \= \kill
  $U$\> $\sim$\> $\mbox{Uniform}\,(0,1)$ \\[1pt]
  $x_1$\> $\leftarrow$\> $\bar L\ +\ U*(\bar R - \bar L)$\\[6pt]
  if $y<f(x_1)$ and $\mbox{Accept}\,(x_1)$ then exit loop \\[6pt]
  if $x_1<x_0$ then $\bar L\ \leftarrow\ x_1$ else $\bar R\ \leftarrow\ x_1$
\end{tabbing}

\caption[]{The `shrinkage' procedure for sampling from the interval.  The
notation $\mbox{Accept}\,(x_1)$ represents a test for whether a point, $x_1$,
that is within $S \cap I$ is an acceptable next state.  If scheme~(1), (2),
or~(3) was used for constructing the interval, all points within $S \cap I$ are 
acceptable.  If the doubling procedure (scheme~(4)) was used,
the point must pass the test of Figure~\ref{fig-test}, below.}\label{fig-samp}

\end{figure}

\begin{figure}[p]

\begin{tabbing}
  \hspace{60pt} \=\+ \hspace{45pt} \= \hspace{13pt} \= \hspace{10pt} \= \kill
  Input:  \> $f$   \> $=$ \> function proportional to the density \\
          \> $x_0$ \> $=$ \> the current point \\
          \> $x_1$ \> $=$ \> the possible next point \\
          \> $y$   \> $=$ \> the vertical level defining the slice \\
          \> $w$   \> $=$ \> estimate of the typical size of a slice \\
          \> $(L,R)\ =\ \mbox{the interval found by the doubling procedure}$ 
          \\[8pt]
  Output: \> whether or not $x_1$ is an acceptable next state
\end{tabbing}\vspace{8pt}

\begin{tabbing}
  \hspace{60pt} \=\+ \kill
  $\hat L\ \leftarrow\ L$,\hspace{6pt} $\hat R\ \leftarrow\ R$ \\
  $D\, \leftarrow\ \mbox{false}$
\end{tabbing}\vspace{8pt}

\begin{tabbing}
\hspace{60pt} \= repeat while $\hat R - \hat L\, >\, 1.1*w$:
\end{tabbing}\vspace{3pt}

\begin{tabbing}
  \hspace{80pt} \=\+ \hspace{15pt} \= \hspace{14pt} \= \kill
  $M$ \> $\leftarrow$\> $(\hat L + \hat R)\, /\, 2$ \\[3pt]
  if \{ $x_0<M$ and $x_1 \ge M$ \} or \{ $x_0 \ge M$ and $x_1<M$ \} then 
    $D\ \leftarrow\ \mbox{true}$ \\[3pt]
  if $x_1<M$ then $\hat R\ \leftarrow\ M$ else $\hat L\ \leftarrow\ M$ \\[3pt]
  if $D$ and $y \ge f(\hat L)$ and $y \ge f(\hat R)$ then \\
    \hspace{20pt}The new point is not acceptable
\end{tabbing}\vspace{8pt}

\begin{tabbing}
\hspace{60pt} \= The new point is acceptable if not rejected in the loop above
\end{tabbing}

\caption[]{The test for whether a new point, $x_1$, that is within $S \cap I$
is an acceptable next state, when the interval was found by the `doubling' 
procedure. The multiplication by 1.1 in the `while' condition guards
against possible round-off error.}\label{fig-test}

\end{figure}

The shrinkage procedure is shown in detail in Figure~\ref{fig-samp}.
Note that each rejected point is used to shrink the interval in such a
way that the current point remains within it.  Since the current point
is always within $A$, the interval used always contains acceptable
points, ensuring that the procedure will terminate.

If the interval was found by scheme~(1), (2), or~(3), the set $A$ is
simply $S \cap I$.  However, if the doubling procedure (scheme~(4))
was used, $A$ may be a smaller subset of $S \cap I$.  This is
illustrated in Figure~\ref{fig-dblp}.  In \ref{fig-dblp}(a), an
interval is found by doubling an initial interval until both ends are
outside the slice.  A different starting point is considered in
\ref{fig-dblp}(b), one which might have been drawn from the interval
found in \ref{fig-dblp}(a).  The doubling procedure terminates earlier
starting from here, so this point is not in $A$.  (Note that $A$ is
here defined conditional on the alignment of the initial interval.)

The $\mbox{Accept}\,(x_1)$ predicate in Figure~\ref{fig-test} tests
whether a point in $S \cap I$ is in $A$ when the doubling procedure
(scheme~(4)) was used. This procedure works backward through the
intervals that the doubling procedure would pass through to arrive at
$I$ when starting from the new point, checking that none of them have
both ends outside the slice, which would lead to earlier termination
of the doubling procedure.  (Note that one needn't check this
explicitly until the intervals differ from those followed from the
current point, a condition tracked with the variable $D$ in the
procedure.)  If the distribution is known to be unimodal, this test
can be omitted, as discussed in section~\ref{sub-unimodal}.

\subsection{Correctness of single-variable slice 
            sampling}\label{sub-correct}\vspace*{-8pt}

To show that single-variable slice sampling is a correct procedure, we
must show that each update leaves the desired distribution invariant.
To guarantee convergence to this distribution, the resulting Markov
chain must also ergodic.  This is not always true, but it is in those
situations (such as when $f(x)>0$ for all $x$) for which one can
easily show that Gibbs sampling is ergodic.  I will not discuss 
the more difficult situations here.

To show invariance, we suppose that the initial state, $x_0$, is
distributed according to $f(x)$.  In step \textit{(a)} of
single-variable slice sampling, a value for $y$ is drawn uniformly
from $(0,f(x))$.  The joint distribution for $x_0$ and $y$ will
therefore be as in equation~(\ref{eq-joint}).  If the subsequent steps
update $x_0$ to $x_1$ in a manner that leaves this joint distribution
invariant, then when we subsequently discard $y$, the resulting
distribution for $x_1$ will be the marginal of this joint
distribution, which is the same as that defined by $f(x)$, as desired.

We therefore need only show that the selection of $x_1$ to follow
$x_0$ in steps~\textit{(b)} and~\textit{(c)} of the single-variable
slice sampling procedure leaves the joint distribution of $x$ and $y$
invariant.  Since these steps do not change $y$, this is the same as
leaving the conditional distribution for $x$ given $y$ invariant, and
this conditional distribution is uniform over \mbox{$S = \{\, x\,:\,y
< f(x)\,\}$}, the slice defined by $y$.  We can show invariance of
this distribution by showing that the updates satisfy detailed
balance, which for a uniform distribution reduces to showing that the
probability density for $x_1$ to be selected as the next state, given
that the current state is $x_0$, is the same as the probability density
for $x_0$ to be the next state, given that $x_1$ is the current state,
for any states $x_0$ and $x_1$ within $S$.

In the process of picking a new state, various intermediate choices
are made randomly.  When the interval is found by the stepping out
procedure of Figure~\ref{fig-step}, the alignment of the initial
interval is randomly chosen, as is the division of the maximum number
of intervals into those used to extend to the left and those used to
extend to the right.  For the doubling procedure of
Figure~\ref{fig-dbl}, the alignment of the initial interval is random
and the decisions whether to extend to the right or to the left are
also made randomly.  When sampling is done using the shrinkage procedure of
Figure~\ref{fig-samp}, zero or more rejected points will be chosen
before the final point.  Let $r$ denote these intermediate random choices.
I will prove that detailed balance holds for the entire procedure by showing
the following stronger result:
\begin{eqnarray}
  \lefteqn{P(\mbox{next state} = x_1,\ \mbox{intermediate choices} = r\ |\ 
    \mbox{current state} = x_0)}\ \ \ \ \nonumber\\
  & = &
  P(\mbox{next state} = x_0,\ \mbox{intermediate choices} = \pi(r)\ |\ 
    \mbox{current state} = x_1)\label{eq-balance}
\end{eqnarray}
where $\pi(r)$ is some one-to-one mapping that has Jacobian one (with
regard to the real-valued variables), which may depend on $x_0$ and $x_1$.  
Integrating over all possible values for $r$ then gives the desired result.

In detail, the mapping $\pi$ used is as follows.  First, if the
interval $I$ is found by the stepping out or doubling procedure,
an intermediate value, $U$, will be generated by the procedure of
Figure~\ref{fig-step} or~\ref{fig-dbl}, and used to define the initial
interval.  We define $\pi$ so that it maps the value $U_0$ chosen when 
the state is $x_0$ to the following $U_1$ when the state is $x_1$:
\begin{eqnarray}
  U_1 & = & \mbox{Frac}\,(U_0+(x_1-x_0)/w)
\end{eqnarray}
where $\mbox{Frac}\,(x) = x-\mbox{Floor}\,(x)$ is the fractional part of $x$. 
This mapping associates values that produce the same alignment
of the initial interval.  Note also that it has Jacobian one.
If the stepping out procedure is used, a value for $J$ is also generated, 
uniformly from the set $\{\,0,\,\ldots,\,m\!-\!1\,\}$.  The mapping 
$\pi$ associates the $J_0$ found when the state is $x_0$ with the 
following $J_1$ when the state is $x_1$:
\begin{eqnarray}
  J_1 & = & J_0\, +\, (x_1/w-U_1)\, -\, (x_0/w-U_0)
\end{eqnarray}
Here, $(x_1/w-U_1)\,-\,(x_0/w-U_0)$ is an integer giving the number
of steps (of size $w$) from the left end of the interval containing $x_0$ to 
the left end of the interval containing $x_1$.  This is the amount by
which we must adjust $J_0$ in order to ensures that if the interval found 
starting from $x_0$ grows to its maximum size, the associated interval found 
starting from $x_1$ will be identical.  Similarly, if the doubling procedure of 
Figure~\ref{fig-dbl} is used, the sequence of random decisions as to
which side of the interval to expand is mapped by $\pi$ to the sequence
of decisions that would cause the interval expanding from $x_1$ to
become identical to the interval expanding from $x_0$ when the latter
first includes $x_1$, and to remain identical through further expansions.  
Note in this respect that there is at most one way that an given final
interval can be obtained by successive doublings from a given initial
interval, and that the alignment of the initial intervals by the 
association of $U_0$ with $U_1$ ensures that doubling starting from $x_1$
can indeed lead to the same interval as found from $x_0$.  Finally, to
complete the definition, $\pi$ maps the sequence of rejected points used
to shrink the interval found from $x_0$ (see Figure~\ref{fig-samp}) 
to the same sequence of points when $x_1$ is the start state.

It remains to show that with this definition of $\pi$,
equation~(\ref{eq-balance}) does indeed hold, for all points $x_0$ and
$x_1$, and all possible intermediate values $r$.  The equation
certainly holds when both sides are zero, so we needn't consider
situations where movement between $x_0$ and $x_1$ is impossible (in
conjunction with the given intermediate values).

Consider first the probability (density) for producing the
intermediate values that define the interval $I$.  For the stepping
out and doubling procedures, the values $U_0$ and $U_1$ (related by
$\pi$) that are generated from $x_0$ and $x_1$ will certainly have the
same probability density, since $U$ is drawn from a uniform
distribution.  Similarly, for the stepping out procedure, the values
$J_0$ and $J_1$ are drawn from a uniform distribution over
$\{\,0,\,\ldots,\,m\!-\!1\,\}$, and hence have the same probability as
long as $J_0$ and $J_1$ are both in this set, which will be true
whenever movement between $x_0$ and $x_1$ is possible.  For the
doubling procedure, a sequence of decisions as to which side to extend
is made, with all sequences of a given length having the same
probability.  Here also, the sequences associated by $\pi$ will have
the same probability, \textit{provided} the same number of doublings
are done starting from $x_0$ as from $x_1$.  This need not be true in
general, but if the sequence from $x_1$ is shorter, the test of
Figure~\ref{fig-test} will eliminate $x_1$ as a possible successor to
$x_0$, and if the sequence from $x_0$ is shorter, $x_1$ will not be a
possible successor because it will be outside the interval $I$ found
from $x_0$.  Both sides of equation~\ref{eq-balance} will therefore be
zero in this situation.

Note next that the intervals found by any of the schemes of
section~\ref{sub-int} will be the same for $x_0$ as for $x_1$, when
the intermediate values chosen are related by $\pi$, assuming a
transition from $x_0$ to $x_1$ is possible.  For the stepping out
procedure, the maximum extent of the intervals will be the same
because of the relationships between $U_0$ and $U_1$ and between $J_0$
and $J_1$.  Furthermore, the actual intervals found by stepping out
(limited by the maximum) must also be the same whenever a transition
between $x_0$ and $x_1$ is possible, since if the interval starting
from $x_0$ has reached $x_1$, expansion of both intervals will
continue in the same direction until the outside of the slice or the
maximum is reached, and likewise in the other direction.  Similarly,
the mapping $\pi$ is defined to be such that if the interval found by
the doubling procedure starting from $x_0$ includes $x_1$, the same
interval would be found from $x_1$, provided the process was not
terminated earlier (by both ends being outside the slice), in which
case $x_1$ is not a possible successor (as it would be rejected by the
procedure of Figure~\ref{fig-test}).  Note also that since the set $A$
is determined by $I$ (for any start state), it too will be the same
for $x_0$ as for $x_1$.

If we sample from this $I$ by simple rejection (scheme~\textit{(i)} in
section~\ref{sub-samp}), the state chosen will be uniformly
distributed over $A$, so the probability of picking $x_0$ will be the
same as that of picking $x_1$.  If we instead use the shrinkage
procedure (scheme~\textit{(ii)}, in Figure~\ref{fig-samp}), we need to
consider as intermediate values the sequence of rejected points that
were used to narrow the interval (recall that under $\pi$ this
sequence is the same for $x_0$ as for $x_1$).  The probability density
for the first of these is clearly the same for both starting points,
since $I$ is the same.  As the interval shrinks, it remains the same
for both $x_0$ and $x_1$, since the rejection decisions (based on $A$)
are the same, and since we need consider only the case where the same
end of the interval is moved to the rejected point (as otherwise a
transition between $x_0$ and $x_1$ in conjunction with these
intermediate values would be impossible).  The probability densities
for later rejected points, and for the final accepted state, are
therefore also the same.

This completes the proof.  Various seemingly reasonable modifications
--- such as changing the doubling procedure of Figure~\ref{fig-dbl} to
not expand the interval on a side that is already outside the slice
--- would undermine the argument of the proof, and hence cannot be
used.  However, some shortcuts are allowed when the distribution is
unimodal, as discussed next.

\subsection{Shortcuts for unimodal 
            distributions}\label{sub-unimodal}\vspace*{-8pt}

Certain shortcuts can be used when the conditional distribution for
the variable being updated is known to be unimodal, or more generally,
when the slice, $S$, is known to consist of a single interval.  For
some values of the auxiliary variable, $S$ may be a single interval
even when the distribution is multimodal, but the effort required to
confirm this probably exceeds the gain from using the shortcuts, so I
will refer only to the unimodal case here.

Two shortcuts apply when the `doubling' procedure is used to find the
interval.  First, for a unimodal distribution, the acceptance test in
Figure~\ref{fig-test} can be omitted, since it will always indicate
that the new point is acceptable.  To see this, note that the
procedure rejects a point when one of the intervals found by doubling
from that starting point has both ends outside the slice, but does not
contain the current point.  Since both the current point and the new
point are inside the slice, this is impossible if the slice consists
of only one interval.

Second, the interval found by the doubling procedure can sometimes be
shrunk at the outset.  The side chosen for extension when the interval
doubles will sometimes be outside the slice already.  When the
distribution is known to be unimodal, it is not possible for such an
extension to contain any points within the slice.  Accordingly, before
sampling is begun, the endpoints of the interval can be set to the
first point in each direction that was found to lie outside the slice.
This may reduce the number of points generated, while having no effect
on the distribution of the point finally chosen.

Finally, if the distribution is known to be unimodal \textit{and} no
limit is imposed on the size of the interval found (ie, $m$ and $p$ in
Figures~\ref{fig-step} and~\ref{fig-dbl} are infinite), the estimate,
$w$, for the typical size of a slice can be set on the basis of past
iterations.  One could, for example, keep a running average of the
distance between the old and new points in past iterations, and use
this (or some suitable multiple) as the estimate $w$.  This is valid
because the distribution of the new point does not depend on $w$ in
this situation, even though $w$ influences how efficiently this new
point is found.  Indeed, when the distribution is known to be
unimodal, one can use any method at all for finding an interval that
contains the current point and has both ends outside the slice, as any
such interval will lead to the new point finally chosen being drawn
uniformly from the slice.

\section{Multivariate slice sampling methods}\label{sec-multi}\vspace*{-10pt}

Rather than sample from a distribution for $x = (x_1,\ldots,x_n)$ by
applying one of the single-variable slice sampling procedures
described above to each $x_i$ in turn, we might try instead to apply
the idea of slice sampling directly to the multivariate distribution.
I will start by describing a straightforward generalization of the
single-variable methods to multivariate distributions, and then
describe a more sophisticated method, which can potentially allow for
adaptation to the local dependencies between variables.

\subsection{Multivariate slice sampling with 
hyperrectangles}\label{sec-hyper}\vspace{-8pt}

We can generalize the single-variable slice sampling methods of
Section~\ref{sec-sing} to methods for performing multivariate updates
by replacing the interval $I=(L,R)$ by an axis-aligned hyperrectangle
$H = \{ x : L_i < x_i < R_i\ \mbox{for all $i=1,\ldots,n$} \}$.  Here,
$L_i$ and $R_i$ define the extent of the hyperrectangle along the axis
for variable $x_i$.

The procedure for finding the next state, $x_1 = (x_{1,1},\ldots,x_{1,n})$,
from the current state, $x_0 = (x_{0,1},\ldots,x_{0,n})$, parallels the 
single-variable procedure:
\begin{enumerate}\vspace*{-8pt}
\item[\textit{a)}] 
          Draw a real value, $y$, uniformly from $(0,f(x_0))$, to define
          the slice \mbox{$S = \{\, x\,:\,y < f(x)\,\}$}.  
\item[\textit{b)}] 
          Find a hyperrectangle, $H = (L_1,R_1)\times\cdots\times(L_n,R_n)$, 
          around $x_0$, which preferably contains at least a big part of the 
          slice.\vspace*{-2pt}
\item[\textit{c)}] 
          Draw the new point, $x_1$, from the part of the
          slice within this hyperrectangle (ie, from $S \cap H$).\vspace*{-8pt}
\end{enumerate}

It would perhaps be ideal for step (b) to set $H$ to the smallest
hyperrectangle containing $S$, but this is unlikely to be feasible.
An easy option when all the variables have bounded range is to set $H$
to be the entire space, but this will often be rather inefficient,
since $S$ is likely to be much smaller.  

In practice, we must usually be content to find an $H$ that contains
the current point, $x_0$, but probably not all of $S$, using width
parameters, $w_i$, for the dimensions of $H$ along each axis.  If we
know nothing about the relative scales of different variables, we
might set all the $w_i$ to a single scale parameter, $w$.  The
simplest way of finding $H$ is to randomly position a hyperrectangle
with these dimensions, uniformly over positions that lead to $H$
containing $x_0$.  This generalizes the random positioning of the
initial interval $I$ for the single-variable slice sampling methods.
The stepping out and doubling procedures used with single-variable
slice sampling do not generalize so easily, however.  The goal of
finding an interval whose endpoints are outside the slice would
generalize to finding a hyperrectangle all of whose vertices are
outside the slice, but since an $n$ dimensional hyperrectangle has
$2^n$ vertices, we would certainly not want to test for this when $n$
is large.  The stepping out procedure seems to be too time consuming
in any case, since one would need to step out in each of the $n$
directions.  The doubling procedure does generalize appropriately, and
one could decide to stop doubling when a randomly-drawn point picked
uniformly from the current hyperrectangle is outside the slice.  This
idea is worth exploring, but here I will consider only the simplest
scheme, which is to use the randomly positioned hyperrectangle without
any expansion, though it is then crucial that the $w_i$ not be
much smaller than they \mbox{should be}.

\begin{figure}[t]

\vspace*{-1pt}

\centerline{\psfig{figure=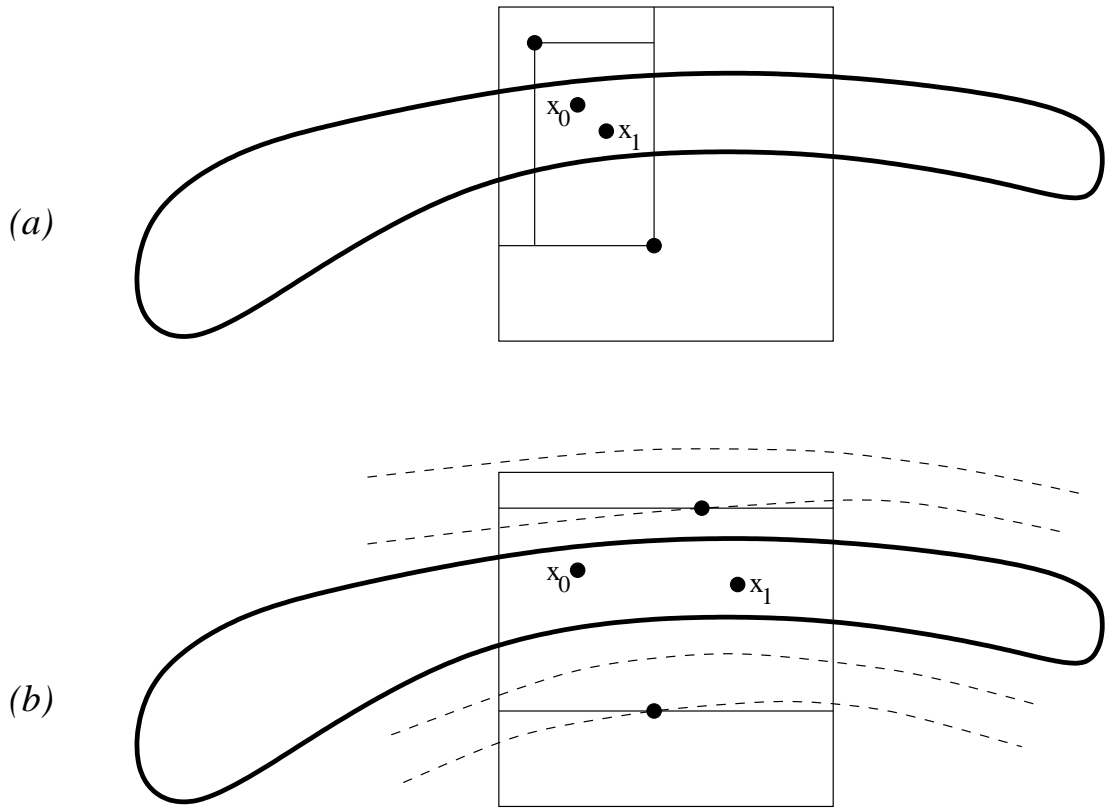}}

\vspace*{4pt}

\caption[]{Multivariate slice sampling with hyperrectangles.  The
heavy line outlines the slice, containing the current point, $x_0$.
The large square is the initial hyperrectangle.  In (a), the
hyperrectangle is shrunk in all directions when the point drawn is
outside the slice, until a new point, $x_1$, inside the slice is
found.  In (b), the hyperrectangle is shrunk along only one axis,
determined from the gradient and the current dimensions of the
hyperrectangle.  The dashed lines are contours of the density
function, indicating the direction of the gradient.}\label{fig-hyper}

\end{figure}

\begin{figure}[p]

\begin{tabbing}
  \hspace{60pt} \=\+ \hspace{45pt} \= \hspace{13pt} \= \hspace{10pt} \= \kill
  Input:  \> $f$   \> $=$ \> function proportional to the density \\
          \> $x_0$ \> $=$ \> the current point, of dimension $n$ \\
          \> $w_i$ \> $=$ \> scale estimates for each variable, $i=1,\ldots,n$
          \\[8pt]
  Output: \> $x_1$ \> $=$ \> the new point
\end{tabbing}\vspace{11pt}

\begin{tabbing}
  \hspace{60pt} \=\+ \hspace{15pt} \= \hspace{14pt} \= \kill
  \textit{Step (a): Find the value of $y$ that defines the slice.} \\[5pt]
  $y$\> $\sim$\> $\mbox{Uniform}\,(0,\,f(x_0))$
\end{tabbing}\vspace{11pt}

\begin{tabbing}
  \hspace{60pt} \=\+ \hspace{15pt} \= \hspace{14pt} \= \kill
  \textit{Step (b): Randomly position the hyperrectangle 
  $H = (L_1,R_1)\times\cdots\times(L_n,R_n)$.} \\[5pt]
  For $i\, =\, 1$ to $n$:
\end{tabbing}\vspace{3pt}
\begin{tabbing}
  \hspace{80pt} \=\+ \hspace{15pt} \= \hspace{14pt} \= \kill
  $U_i$\> $\sim$\> $\mbox{Uniform}\,(0,1)$ \\
  $L_i$\> $\leftarrow$\> $x_{0,i}\ -\ w_i*U_i$ \\
  $R_i$\> $\leftarrow$\> $L_i\ +\ w_i$
\end{tabbing}\vspace{11pt}

\begin{tabbing}
  \hspace{60pt} \=\+ \hspace{15pt} \= \hspace{14pt} \= \kill
  \textit{Step (c): Sample from $H$, shrinking when points are rejected.}\\[5pt]
  Repeat: \\[3pt]
  \hspace{20pt}For $i\, =\, 1$ to $n$:
\end{tabbing}\vspace{3pt}
\begin{tabbing}
  \hspace{100pt} \=\+ \hspace{20pt} \= \hspace{14pt} \= \kill
  $U_i$\> $\sim$\> $\mbox{Uniform}\,(0,1)$ \\
  $x_{1,i}$\> $\leftarrow$\> $L_i\ +\ U_i*(R_i-L_i)$
\end{tabbing}\vspace{11pt}
\begin{tabbing}
  \hspace{80pt} \=\+ \hspace{15pt} \= \hspace{14pt} \= \kill
  if $y<f(x_1)$ then exit loop \\[6pt]
  For $i\, =\, 1$ to $n$: \\
  \hspace{20pt}if $x_{1,i}<x_{0,i}$ then 
  $L_i\ \leftarrow\ x_{1,i}$ else $R_i\ \leftarrow\ x_{1,i}$
\end{tabbing}

\caption[]{A simple multivariate slice sampling procedure, with randomly
positioned hyperrectangle, and shrinkage in all directions.}\label{fig-simph}

\end{figure}

The shrinkage procedure of Figure~\ref{fig-samp} generalizes easily to
multiple dimensions --- the hyperrectangle can simply be shrunk
independently along each axis.  Combining this with simple random
positioning of $H$, one gets the multivariate slice sampling method
shown in Figure~\ref{fig-hyper}(a), and given in detail in
Figure~\ref{fig-simph}.  The validity of this method can be proven in
the same way as was done for single-variable slice sampling in
Section~\ref{sub-correct}.

Although this simple multivariate slice sampling method is easily
implemented, and will often be reasonably efficient, in one respect it
works less well than applying single-variable slice sampling to each
variable in turn.  When each variable is updated separately, the
interval for that variable will be shrunk only as far as needed in
order to obtain a new value within the slice.  The amount of shrinkage
can be different for different variables.  In contrast, the procedure
of Figure~\ref{fig-simph} shrinks all dimensions of the hyperrectangle
until a point inside the slice is found, even though the probability
density may not vary rapidly in some of these dimensions, making
shrinkage in these directions unnecessary (and undesirable).

One way to try to avoid this problem is illustrated in
Figure~\ref{fig-hyper}(b).  Rather than shrink all dimensions of the
hyperrectangle when the last point chosen was outside the slice, we
can instead shrink along only one axis, basing the choice on the
gradient of $\log f(x)$, evaluated at the last point.  Specifically,
only the axis corresponding to variable $x_i$ is shrink, where $i$
maximizes the following product:\vspace*{-6pt}
\begin{eqnarray}
  (R_i \,-\, L_i)\ |G_i|
\label{eq-hyperprod}\end{eqnarray}
where $G$ is the gradient of $\log f(x)$ at the last point chosen.
By multiplying the magnitude of component $i$ of the gradient by the 
width of the hyperrectangle in this direction, we get an estimate of the 
amount by which $\log f(x)$ changes along axis $i$.  The axis for which
this change is thought to be largest is likely to be the best one to shrink
in order to eliminate points outside the slice.
Unfortunately, if this decision were based as well on whether the sign of the
gradient indicates that $\log f(x)$ is increasing or decreasing as we
move toward the current point, $x_0$, the shrinkage decision might be
different if we were to shrink from the final accepted point, $x_1$,
which would invalidate the method (unless we somehow avoided or rejected
such points).

Many more elaborate schemes along these lines are possible.  For
instance, we might shrink along all axes for which the product
(\ref{eq-hyperprod}) is greater than some threshold.  A good scheme
might preserve the ability of single-variable slice sampling to adapt
differently for different variables, while keeping the advantages that
simultaneous updates may sometimes have (eg, in producing an ergodic
chain when there are tight dependencies between variables).

More ambitiously, we might hope that a multivariate slice sampler
could adapt to the dependencies between variables, not just to their
different scales.  This will require that we go beyond axis-aligned
hyperrectangles, as is done in the next section.

\subsection{A framework for adaptive multivariate slice
sampling}\label{sec-multiadapt}\vspace{-8pt}

We would like a more general framework by which trial points outside
the slice that were previously rejected can be used to guide the
selection of future trial points.  In contrast to schemes based on
hyperrectangles, we would like future trial points to potentially come
from distributions that take account of the dependencies between
variables.  The scheme I present here achieves this by laying down a
trail of `crumbs' that guide the selection of future trial points,
leading eventually to a point inside the slice.  A crumb can be
anything --- eg, a discrete value, a real number, a vector, a
hyperrectangle --- but the method is perhaps most easily visualized
when crumbs are points in the state space being \mbox{sampled from}.

As with the previous slice sampling schemes, we start by chosing a
value $y$ uniformly between zero and $f(x_0)$, where $x_0$ is the
current point.  A crumb, $c_1$, is then draw at random from some
distribution with density (or probability mass) function
$g_1(c;\,x_0,\,y)$.  Note that this distribution may depend on both
the current point, $x_0$, and on the value of $y$ that defines the
slice.  A first trial point, $x^*_1$, is then drawn from the
distribution with density $h_1(x^*;\,y,\,c_1) =
g_1(c_1;\,x^*,\,y)\,/\,Z_1(y,\,c_1)$, where $Z_1(y,\,c_1)=\int
g_1(c_1;\,x^*,\,y)\, dx^*$ is the appropriate normalizing constant.
One can view $x^*_1$ as being drawn from a pseudo-posterior
distribution, based on a uniform prior, and the ``data'' that the
first crumb was $c_1$.  If $x^*_1$ is inside the slice, we set the new
point, $x_1$, to $x^*_1$, and are finished.  Otherwise, a second
crumb, $c_2$, is drawn from some distribution
$g_2(c;\,x_0,\,y,\,c_1,\,x^*_1)$, which may depend on the previous
crumb and the previous trial point, as well as $x_0$ and $y$.  The
second trial point is then drawn from the pseudo-posterior
distribution based on the ``data'' $c_1$ and $c_2$ --- that is,
$x^*_2$ is drawn from
\begin{eqnarray}
  h_2(x^*;\,y,\,c_1,\,x^*_1,\,c_2) & = & 
  g_1(c_1;\,x^*,\,y)\,g_2(c_2;\,x^*,\,y,\,c_1,\,x^*_1)
   \,/\,Z_2(y,\,c_1,\,x^*_1,\,c_2)
\end{eqnarray}
where $Z_2(y,\,c_1,\,x^*_1,\,c_2) = \int g_1(c_1;\,x^*,\,y)\,
g_2(c_2;\,x^*,\,y,\,c_1,\,x^*_1)\,dx^*$.
If $x^*_2$ is inside the slice, it becomes the new state.  Otherwise,
we draw a third crumb, from a distribution that may depend on the
current state, the value defining the slice, the previous crumbs,
and the previous trial points, generate a third trial point using
this and the previous crumbs, and so forth until a trial point
lying within the slice is found.

To show that this procedure leaves the distribution with density
$f(x)/Z$ invariant, it suffices to show that it separately satisfies
detailed balance with respect to transitions that occur in conjunction
with any given number of crumbs being drawn.  In the case, for instance,
of transitions involving two crumbs, we can show this by showing the
stronger property that for any $x^*_1$ that is not in the slice defined
by $y$ and any $x^*_2$ that is in this slice, the following will hold:
\begin{eqnarray}
  P(x_0)\ P(y,\,c_1,\,x^*_1,\,c_2,\,x^*_2\ |\ x_0) & = &
  P(x^*_2)\ P(y,\,c_1,\,x^*_1,\,c_2,\,x_0\ |\ x^*_2)
\label{eq-crumbal}\end{eqnarray}
Here, $P(x_0)$ and $P(x^*_2)$ are the probability densities for the
current point and the point that will become the new point (which are
proportional to $f(x)$).  The conditional probabilities above are the
densities for the given sequence of values being chosen during the
procedure, given that the current point is the one conditioned on.
The left side of equation~(\ref{eq-crumbal}) can be written as
follows:
\begin{eqnarray*}
  \lefteqn{P(x_0)\cdot P(y\ |\ x_0)\cdot P(c_1\ |\ x_0,\,y)\cdot
  P(x^*_1\ |\ y,\,c_1) \cdot P(c_2\ |\ x_0,\,y,\,c_1,\,x^*_1)\cdot
  P(x^*_2\ |\ y,\,c_1,\,x^*_1,\,c_2)}\ \ \ \ \ \ \\[3pt]
  & = & [f(x_0)/Z]\ \cdot\ [1/f(x_0)] 
  \ \cdot\ 
  g_1(c_1;\,x_0,\,y) \ \cdot\ [g_1(c_1;\,x^*_1,\,y)\,/\,Z_1(y,\,c_1)] \\
  & & \ \ \ \ \ \cdot\ 
  g_2(c_2;\,x_0,\,y,\,c_1,\,x^*_1)
   \ \cdot\ [g_1(c_1;\,x^*_2,\,y)\,g_2(c_2;\,x^*_2,\,y,\,c_1,\,x^*_1)
     \,/\,Z_2(y,\,c_1,\,x^*_1,\,c_2)]
\end{eqnarray*}
The right side is
\begin{eqnarray*}
  \lefteqn{P(x^*_2)\cdot P(y\ |\ x^*_2)\cdot P(c_1\ |\ x^*_2,\,y)\cdot
  P(x^*_1\ |\ y,\,c_1) \cdot P(c_2\ |\ x^*_2,\,y,\,c_1,\,x^*_1)\cdot
  P(x_0\ |\ y,\,c_1,\,x^*_1,\,c_2)}\ \ \ \ \ \ \\[3pt]
  & = & [f(x^*_2)/Z] \ \cdot\ [1/f(x^*_2)] 
  \ \cdot\ 
  g_1(c_1;\,x^*_2,\,y) \ \cdot\ [g_1(c_1;\,x^*_1,\,y)\,/\,Z_1(y,\,c_1)] \\
  & & \ \ \ \ \ \cdot\ 
  g_2(c_2;\,x^*_2,\,y,\,c_1,\,x^*_1)
   \ \cdot\ [g_1(c_1;\,x_0,\,y)\,g_2(c_2;\,x_0,\,y,\,c_1,\,x^*_1)
     \,/\,Z_2(y,\,c_1,\,x^*_1,\,c_2)]
\end{eqnarray*}
These are readily seen to be equal, as is true in general for transitions
involving any number of crumbs.

The hyperrectangle methods of Section~\ref{sec-hyper} can be viewed in
this framework.  The randomly placed initial hyperrectangle is the
first crumb.  The first trial point is chosen from those points that
could produce this initial hyperrectangle, which is simply the set of
points within the hyperrectangle.  The second and later crumbs are the
shrunken hyperrectangles.  Conditional on the current point, the
previous crumb (ie, the previous hyperrectangle), and the previous
trial point, these have degenerate distributions, concentrated on a
single hyperrectangle.  The possible corresponding trial points are
the points within the shrunken hyperrectangle.

By using different sorts of crumbs, and different distributions for
them, a huge variety of methods could be constructed within this
framework.  I will here only briefly discuss methods in which the
crumbs are points in the state space, and have multivariate Gaussian
distributions.  The distributions of the trial points given the crumbs
will then also be multivariate Gaussians.

In the simplest method of this sort, every $g_i$ is Gaussian with mean
$x_0$ and covariance matrix $\sigma^2 I$, for some fixed $\sigma^2$.
The distribution, $h_i$, for $x^*_i$ will then be Gaussian with mean
$\bar c_i = (c_1+\ldots+c_i)/i$ and covariance matrix $(\sigma^2/i)
I$.  As more and more trial points are generated, they will come from
narrower and narrower distributions, which will be concentrated closer
and closer to the current point (since $\bar c_i$ will approach
$x_0$).  This is analogous to shrinkage in the hyperrectangle method.
In practice, it would probably be desirable to let $\sigma^2_i$
decrease with $i$ (perhaps exponentially), so that the trial points
would be forced closer to $x_0$ more quickly.  Alternatively, one
might look at $f(x^*_{i-1})/y$ in order to estimate what value for
$\sigma_i$ would produce a distribution for the next trial point,
$x^*_i$, that is likely to lie within the slice.

More generally, $g_i$ could be a multivariate Gaussian with mean $x_0$
and some covariance matrix $\Sigma_i$, which may depend on the value of
$y$, the previous crumbs, and the previous trial points.  In
particular, $\Sigma_i$ could depend on the gradients of $f(x^*_j)$ for
$j<i$, which provide information on what Gaussian distribution would
be a good local approximation to $f(x)$.  The distribution, $h_i$, for
trial point $x^*_i$ will then have covariance $\Sigma^*_i = [\Sigma^{-1}_1 +
\cdots + \Sigma^{-1}_{i}]^{-1}$ and mean $\bar c_i = \Sigma^*_i \,
[\Sigma^{-1}_1 c_1 + \cdots + \Sigma^{-1}_i c_i]$.

When $x$ is of only moderate dimensionality, explicitly performing
operations involving these covariance matrices would be tolerable, and
a wide variety of ways for producing $\Sigma_i$ would be feasible.
For higher-dimensional problems, such operations would need to be
avoided, as is done in an optimization context with the conjugate
gradient and other related methods.  Further research is therefore
needed in order to fully exploit the potential of this promising
framework for adaptation, and to compare it with methods based on the
`delayed rejection' (also called `splitting rejection') framework of
Tierney and Mira (Mira 1998, Chapter~5; Tierney and Mira 1999).

\section{Overrelaxed slice sampling}\label{sec-over}\vspace*{-8pt}

\begin{figure}[t]
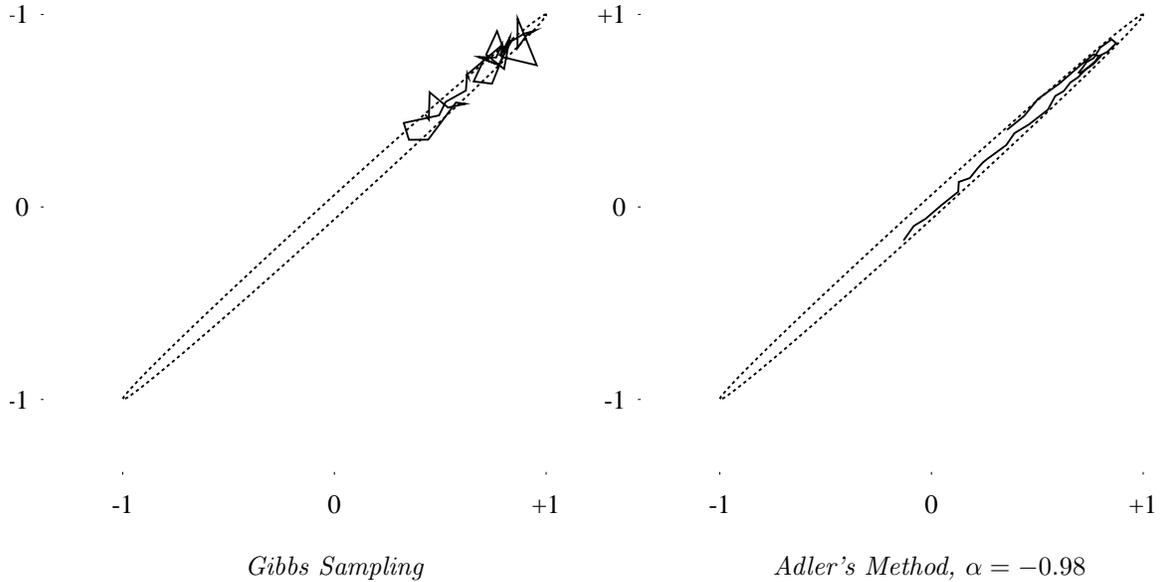


\vspace*{2.7in}
\includegraphics{gibbs-ellipse.ps}
\includegraphics{over-ellipse.ps}

\vspace{12pt}\hspace{10pt}%
\makebox[225pt]{\small \em Gibbs Sampling}%
\makebox[225pt]{\small \em Adler's Method, $\alpha=-0.98$}\hspace*{-30pt}

\vspace*{0.3in}
\hspace{0pt}\parbox{177pt}{\caption[]{
Gibbs sampling and Adler's overrelaxation method applied
to a bivariate Gaussian with correlation 0.998 (whose
one-standard-deviation contour is plotted).  The top left shows the
progress of 40 Gibbs sampling iterations (each consisting of one
update for each variable).  The top right shows 40 overrelaxed iterations,
with $\alpha=-0.98$.  The close-up on the right shows how successive
overrelaxed updates operate to avoid a random walk.
}\label{fig-biover}}
\vspace*{0.6in}

\includegraphics{closeup.ps}
\vspace*{-0.3in}

\end{figure}

When the updates used do not account for the dependencies between
variables, many updates will be needed to move from one part of the
distribution to another. Sampling efficiency can be improved in this
context by suppressing the random walk behaviour characteristic of
simple schemes such as Gibbs sampling.  One way of achieving this is
by using `overrelaxed' updates.  Like Gibbs sampling, overrelaxation
methods update each variable in turn, but rather than drawing a new
value for a variable from its conditional distribution independently
of the current value, the new value is instead chosen to be on the
opposite side of the mode from the current value.  In Adler's (1981)
scheme, applicable when the conditional distributions are Gaussian,
the new value for variable $i$ is 
\begin{eqnarray}
  x_i^{\prime} & = & \mu_i \,+\, \alpha\,(x_i - \mu_i)
                     \,+\, \sigma_i\,(1-\alpha^2)^{1/2}\,n
\end{eqnarray}
where $\mu_i$ and $\sigma_i$ are the conditional mean and standard deviation
of variable $i$, $n$ is a Gaussian random variate with mean zero and variance
one, and $\alpha$ is a parameter slightly greater than $-1$.  This method
is analysed and discussed by Barone and Frigessi (1990) and by Green and Han
(1992), though these discussions fail in some respects to properly elucidate 
the true benefits and limitations of overrelaxation (Neal 1998).  The crucial
ability of overrelaxation to (sometimes) suppress random walks is illustrated
for a bivariate Gaussian distribution in Figure~\ref{fig-biover}.

Various attempts have been made to produce overrelaxation schemes that
can be used when the conditional distributions are not Gaussian.  I
have reviewed several such schemes, and introduced one of my own (Neal
1998).  The concept of overrelaxation seems to apply only when the
conditional distributions are unimodal, so we may assume that this is
usually the case, though we would like the method to at least remain
valid (ie, leave the desired distribution invariant) even if this
assumption turns out to be false.  To obtain the full benefits of
overrelaxation, it is important that almost every update be
overrelaxed, with few or no `rejections' that leave the state
unchanged, as such rejections re-introduce an undesirable random walk
aspect to the motion through state space (Neal 1998).

\begin{figure}[t]

\centerline{\psfig{figure=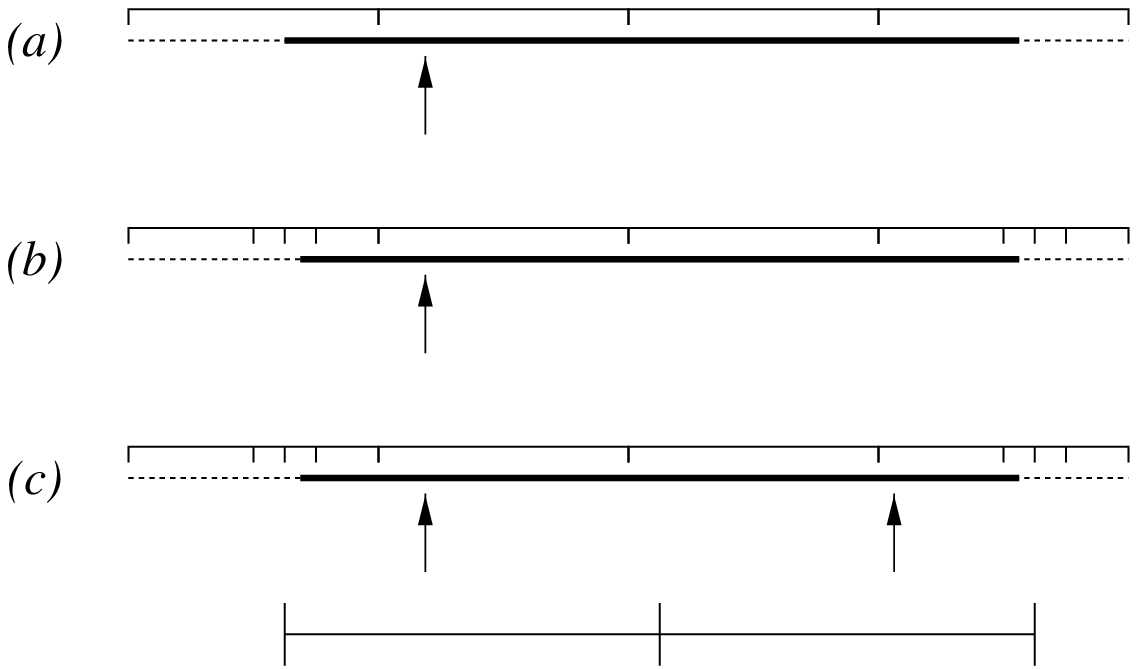}}

\hspace{1.76in} $\hat L$ 
\hspace{1.13in} $\displaystyle {\hat L+\hat R \over 2}$  
\hspace{1.13in} $\hat R$

\caption[]{Overrelaxation using the stepping out procedure and bisection.
In \textit{(a)}, an interval with both ends outside the slice is found 
by stepping out from the current point, as was illustrated in 
Figure~\ref{fig-sing}(b).
In \textit{(b)}, the endpoints of the slice are located more accurately
using bisection.  In \textit{(c)}, a candidate point is found by 
flipping through the point half-way between the approximations to the 
endpoints.  In this case, the candidate point is accepted, since it is within 
the slice, and within the orginal interval (prior to bisection).
}\label{fig-overp}

\end{figure}

In this section, I will show how overrelaxation can be done using
slice sampling.  Many schemes for overrelaxed slice sampling are
possible, but I will describe only one in detail, based on the stepping
out procedure and on bisection.  This scheme is illustrated in
Figure~\ref{fig-overp}, and given in detail in Figure~\ref{fig-over}.

\begin{figure}[p]

\begin{tabbing}
  \hspace{60pt} \=\+ \hspace{45pt} \= \hspace{13pt} \= \hspace{10pt} \= \kill
  Input:  \> $f$   \> $=$ \> function proportional to the density \\
          \> $x_0$ \> $=$ \> the current point \\
          \> $y$   \> $=$ \> the vertical level defining the slice \\
          \> $w$   \> $=$ \> estimate of the typical size of a slice \\
          \> $a$   \> $=$ \> integer limiting endpoint accuracy to $2^{-a}\,w$\\
          \> $(L,R)\ =\ \mbox{interval found by the stepping out procedure}$ 
          \\[8pt]
  Output: \> $x_1$ \> $=$ \> the new point
\end{tabbing}\vspace{8pt}

\begin{tabbing}
  \hspace{60pt} \=\+ \hspace{15pt} \= \hspace{14pt} \= \kill
  $\bar L\ \leftarrow\ L$,\hspace{6pt} $\bar R\ \leftarrow\ R$ \\
  $\bar w\ \leftarrow\ w$,\hspace{6pt} $\bar a\ \leftarrow\ a$ \\[8pt]
  \textit{When the interval is only of size $w$, the following section
          will narrow it} \\
  \textit{until the mid-point is inside the slice (or
          the accuracy limit is reached).} \\[8pt]
  if $R-L\,<\,1.1*w$ then
\end{tabbing}\vspace{3pt}

\begin{tabbing}
  \hspace{80pt} \=\+ \hspace{15pt} \= \hspace{14pt} \= \kill
  repeat:
\end{tabbing}\vspace{3pt}

\begin{tabbing}
  \hspace{100pt} \=\+ \hspace{15pt} \= \hspace{14pt} \= \kill
  $M$\> $\leftarrow$\> $(\bar L + \bar R)\,/\,2$ \\[3pt]
  if $\bar a\,=\,0$ or $y\,<\,f(M)$ then exit loop \\[3pt]
  if $x_0>M$ then $\bar L\ \leftarrow\ M$ else $\bar R\ \leftarrow\ M$ \\[3pt]
  $\bar a$\> $\leftarrow$\> $\bar a - 1$ \\
  $\bar w$\> $\leftarrow$\> $\bar w\,/\,2$
\end{tabbing}\vspace{8pt}

\begin{tabbing}
  \hspace{60pt} \=\+ \hspace{15pt} \= \hspace{14pt} \= \kill
  \textit{Endpoint locations are now refined by bisection, to the
          specified accuracy.}\\[8pt]
  $\hat L\ \leftarrow\ \bar L$,\hspace{6pt} $\hat R\ \leftarrow\ \bar R$ \\[3pt]
  repeat while $\bar a\, > \, 0$:
\end{tabbing}\vspace{3pt}

\begin{tabbing}
  \hspace{80pt} \=\+ \hspace{15pt} \= \hspace{14pt} \= \kill
  $\bar a$\> $\leftarrow$\> $\bar a - 1$ \\
  $\bar w$\> $\leftarrow$\> $\bar w\,/\,2$ \\[3pt]
  if $y\,\ge\,f(\hat L+\bar w)$ then $\hat L\ \leftarrow\ \hat L+\bar w$ \\
  if $y\,\ge\,f(\hat R-\bar w)$ then $\hat R\ \leftarrow\ \hat R-\bar w$ 
\end{tabbing}\vspace{8pt}

\begin{tabbing}
  \hspace{60pt} \=\+ \hspace{15pt} \= \hspace{14pt} \= \kill
 \textit{A candidate point is found by flipping from the current point to the}\\
  \textit{opposite side of $(\hat L, \hat R)$.  It is then tested for 
          acceptability.}
  \\[8pt]
  $x_1$\>$\leftarrow$\> $\hat L\ +\ \hat R\ -\ x_0$ \\[3pt]
  if $x_1<\bar L$ or $x_1>\bar R$ or $y\,\ge\,f(x_1)$ then 
\end{tabbing}\vspace{3pt}

\begin{tabbing}
  \hspace{80pt} \=\+ \hspace{15pt} \= \hspace{14pt} \= \kill
  $x_1$\> $\leftarrow$\> $x_0$
\end{tabbing}

\caption[]{The overrelaxation procedure using bisection.  It is assumed 
that the interval $(L,R)$ was found by the stepping out procedure,
with a stepsize of $w$.  
}\label{fig-over}

\end{figure}

To begin, we apply the stepping out procedure of Figure~\ref{fig-step}
to find an interval around the current point.  Normally, we would
apply this procedure with the maximum size of the interval ($m$) set
to infinity, or to some large value, since a proper overrelaxation
operation requires that the entire slice be found, but the scheme
remains valid for any $m$.

If the stepping out procedure found an interval around the slice that
is bigger than the initial interval, then the two outermost steps will
serve to locate the endpoints of the slice to within an interval of
size $w$.  (Here, we assume that the slice consists of a single
interval, as it will if the distribution is unimodal.)  We then locate
the endpoints more precisely using a bisection procedure.  For each
endpoint, we test whether the mid-point of the interval within which
it is located is inside or outside the slice, and shrink this interval
appropriately to narrow the location of the endpoint.  This is
repeated $a$ times, after which each endpoint will be known to lie
within an interval of size $2^{-a}\,w$.

If the stepping out procedure found that the initial interval (of size
$w$) already had both ends outside the slice, then before doing any
bisection, we narrow this interval, by shrinking it in half repeatedly
until its mid-point is within the slice.  We then use bisection as
above to locate the endpoints to within an interval of size $2^{-a}\,w$.

Once the locations of the endpoints have been narrowed down, we can 
approximate the
entire slice by the interval $(\hat L, \hat R)$, formed from the outer
bounds on the endpoint locations.  To do an overrelaxed update, we
flip from the current point, $x_0$, to a new point, $x_1$, that is
the same distance as the current point from the middle of this 
interval, but on the opposite side.  That is, we let\vspace{-3pt}
\begin{eqnarray}
  x_1 \ \ =\ \ 
  {\hat L + \hat R \over 2}\ -\ \left( x_0 - {\hat L + \hat R \over 2} \right)
  \ \ =\ \ \hat L \ +\ \hat R\ -\ x_0
\end{eqnarray}

We must sometimes reject this candidate point, in which case the new
point is the same as the current point.  First of all, we must reject
$x_1$ if it lies outside the interval, $(\bar L, \bar R)$, that had
been found prior to bisection, since the interval found from $x_1$
would then be different, and detailed balance would not hold.
However, this situation cannot arise when the distribution is
unimodal. Secondly, we must reject $x_1$ if it lies outside the slice.
This can easily happen for a multimodal distribution, and can happen
even for a unimodal distribution when the endpoints of the slice have
not been located exactly.  However, the probability of rejection for a
unimodal distribution can be reduced to as low a level as desired, at
moderate cost, by locating the endpoints more precisely using more
iterations of bisection.

The correctness of this procedure can be seen using arguments similar
to those of section~\ref{sub-correct}.  The interval before bisection
can be found by the doubling procedure instead of stepping out,
provided the point found is rejected if it fails the acceptance test
of Figure~\ref{fig-test}.  However, rejection for this reason will not
occur in the presumably typical case of a unimodal distribution.

One could use many methods other than bisection to narrow down the
locations of the endpoints before overrelaxing.  If the derivative of
$f(x)$ can easily be calculated, one could use Newton iteration, whose
rapid convergence would often allow the endpoints to be calculated to
machine precision in a few iterations.  For unimodal distributions,
such exact calculations would eliminate the possibility of rejection,
and would also make the final result be independent of the way the
interval containing the slice was found, thereby allowing use of
retrospective methods for tuning the procedure for finding this
interval.

To obtain a full sampling scheme, overrelaxed updates of this sort
would be applied to each variable in turn, in a fixed order, for a
number of cycles, after which a normal slice sampling update would be
done.  Alternatively, each update could be done normally with some
small probability.  A Markov chain consisting solely of overrelaxed
updates might not be ergodic, and might in any case suppress random
walks for too long.  The frequency of normal updates is a tuning
parameter, analogous to the choice of $\alpha$ in Adler's
overrelaxation method, and would ideally be set so that the Markov
chain moves systematically, rather than in a random walk, for long
enough that it traverses a distance comparable to the largest
dimension of the multivariate distribution, but for no longer than
this.  To keep from doing a random walk for around $k$ steps, one
would do every $k$'th update normally, and also arrange for the
rejection rate for the overrelaxed updates to be less than $1/k$.

\section{Reflective slice sampling}\label{sec-reflect}\vspace*{-8pt}

Multivariate slice sampling methods can also be designed to suppress
random walks.  In this section I describe methods that `reflect' off
the boundaries of the slice.  Such movement with reflection can be
seen as a specialization to uniform distributions of the Hamiltonian
dynamics that forms the basis for Hybrid Monte Carlo (Duane,
\textit{et al} 1987).

As before, suppose we wish to sample from a distribution over $\Re^n$,
defined by a function $f(x)$ that is proportional to the probability
density, and which we here assume is differentiable.  We must be able
to calculate both $f(x)$ and its gradient (or equivalently, the value
and gradient of $\log f(x)$).  In each iteration of the Markov chain,
we will draw a value for an auxiliary variable, $y$, uniformly from
$(0,f(x))$, thereby defining an $n$-dimensional slice \mbox{$S = \{\,
x\,:\,y < f(x)\,\}$}.  We will also introduce $n$ additional
`momentum' variables, written as a vector $p$, which serve to indicate
the current direction and speed of motion through state space.  At the
start of each iteration, we pick a value for $p$, independently of
$x$, from some rotationally symmetric distribution, typically Gaussian
with mean zero and identity covariance matrix.

Once $y$ and $p$ have been drawn, we repeatedly update $x$ by stepping
in the direction of $p$.  After some predetermined number of steps, we
take the final value of $x$ as our new state (provided it is
acceptable).  In each step, we try to set \mbox{$x^{\prime} = x +
wp$}, for some scale parameter $w$ that determines the average step
size.  However, if the resulting $x^{\prime}$ is outside the slice $S$
(ie, $y \ge f(x^{\prime})$), we must somehow bring try to bring it
back inside.  The schemes considered here all do this by some form of
reflection, but differ in the exact procedure used.

\begin{figure}[p]

\vspace*{-1pt}

\centerline{\psfig{figure=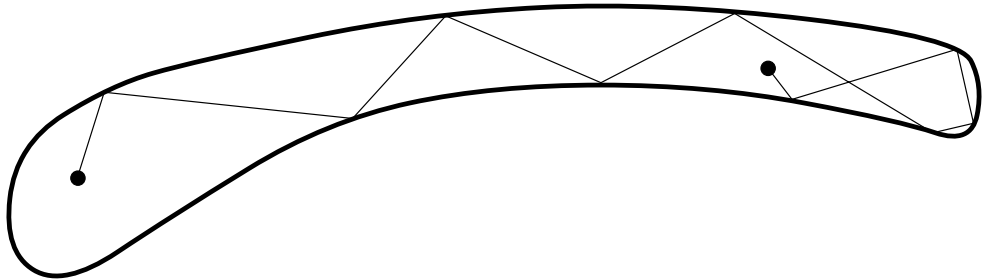}}

\caption[]{Moving around a two-dimensional slice by reflection from the
exact boundaries.}\label{fig-reflect}

\end{figure}

\begin{figure}[p]

\vspace*{-1pt}

\centerline{\psfig{figure=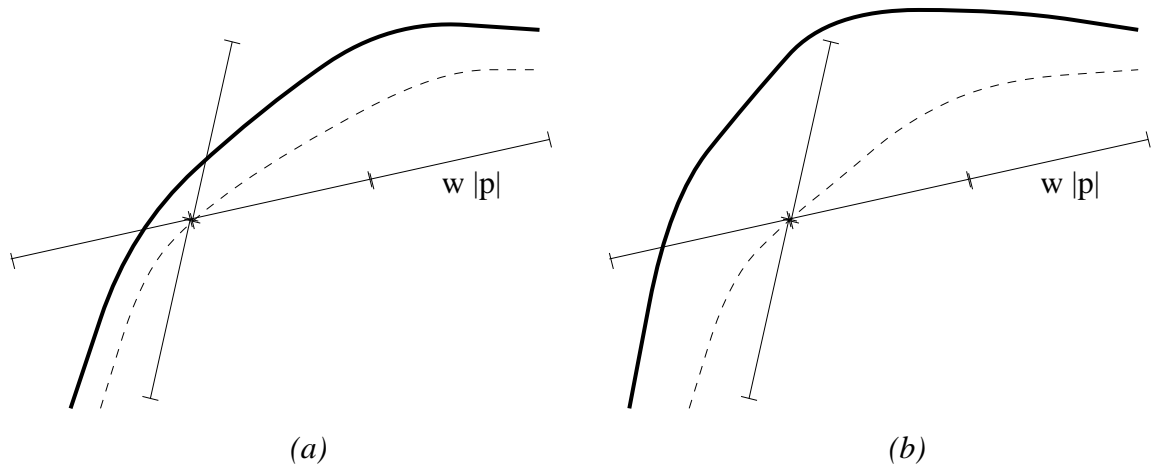,width=6in}}

\caption[]{Reflection from an inside point.  The trajectories here go in steps
of size $w|p|$, starting from the top right, until a point outside
the slice is reached, when a reflection is attempted based
on the inner contour shown.  In \textit{(a)}, the reflection is
is sucessful; in \textit{(b)}, it must be rejected, since the reverse
trajectory would not reflect at this point.}\label{fig-inside}

\end{figure}

\begin{figure}[p]

\vspace*{-1pt}

\centerline{\psfig{figure=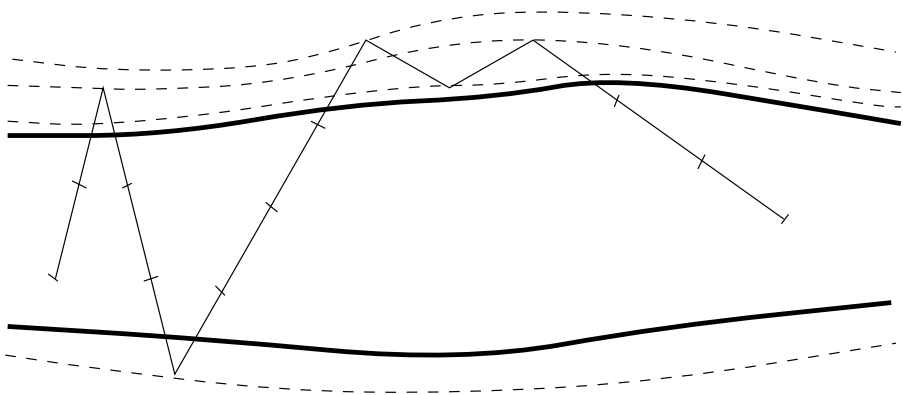}}

\caption[]{Reflection from outside points. Starting from the left, two
reflections based on outside contours lead back inside the slice after
the next step.  The step after the third reflection is still outside
the slice, so further reflections must be done.  In this case, the
trajectory eventually returns to the slice, and its endpoint would
therefore be accepted.}\label{fig-outside}

\end{figure}

Ideally, we would reflect from the exact point at which
movement in the direction of $p$ first takes us outside the slice.
This reflection operation modifies $p$, after which motion continues
in the new direction, until we again encounter the boundary of the slice.
When we hit the boundary at a point where the gradient of $f(x)$ 
is $g$, reflection will change $p$ as follows:\vspace*{-6pt}
\begin{eqnarray} 
 p^{\prime} & = & p\ -\ 2\, g \, {p \cdot g \over |g|^2}
\label{eq-reflect}
\end{eqnarray}
This ideal reflection scheme is illustrated for a two-dimensional
slice in Figure~\ref{fig-reflect}.  Using the fact that the 
reflection transformation above has Jacobian one and is its own inverse,
one can show that movement with reflection for some pre-determined duration 
leaves invariant the joint distribution of $x$ (uniform within the slice) and 
$p$ (rotationally symmetric, independent of $x$), so this
way of sampling is valid, with no need for an acceptance test.  One can also 
see from the figure how such motion can proceed consistently in one 
direction (until the end of the slice is reached), rather than in a random walk.

Ideal reflection is difficult to implement, however, as it requires
precise calculation of where the current path intersects the
boundary of the slice.  Finding this point analytically might
sometimes be possible, or we might try to solve for it numerically,
but if the slice is not known to be convex, it may be difficult even
to determine with certainty that an intersection point that has been
found is in fact the first one that would be encountered.  Rather than
attempt such exact calculations, we can instead employ one of two
approximate schemes, based on `inside' or `outside' reflection,
although the trajectories these schemes produce must sometimes be
rejected.

When stepping from $x$ to $x^{\prime} = x + wp$ takes us outside the
slice, we can try to reflect from the last inside point, $x$, instead
of from the exact point where the path intersects the boundary, using
the gradient of $f(x)$ at this inside point.  The process is
illustrated in Figure~\ref{fig-inside}.  However, for this method to
be valid, we must check that the reverse trajectory would also reflect
at this point, by verifying that a step in the direction opposite to
our new heading would take us outside the slice.  If this is not so,
we must either reject the entire trajectory of which this reflection
step forms a part, or alternatively, set $p$ and $x$ so that we
retrace the path taken to this point (starting at the inside point
where the reflection failed).

Alternatively, when we step outside the slice, we can try to reflect
from the outside point, $x^{\prime}$, based on the gradient at that
point.  A trajectory with several such reflections is shown in
Figure~\ref{fig-outside}.  After performing a pre-determined number of
steps, we accept the trajectory if the final point is inside the
slice.  Note that for this method to be valid, one must reflect
\textit{whenever} the current point is outside the slice, even if this
leads one away from the slice rather than toward it.  This will
sometimes result in the trajectory never returning to the slice, and
hence being rejected, but other times, as in the figure, it does
return eventually.

Many variations on these procedures are possible.  Above, it was
assumed that values for $y$ and $p$ are are randomly drawn at the
beginning of a trajectory, and then kept the same for many steps
(apart from the changes to $p$ when reflections occur).  When using
inside reflection, we might instead pick a new value for $y$ more
often, perhaps before every step, and we might also partially update
$p$, as is done in Horowitz's (1991) variation on Hybrid Monte Carlo.
When using outside reflection, the acceptance rate can be increased by
terminating the trajectory when either some pre-set maximum number of
steps have been taken, \textit{or} some pre-set number of steps have
ended inside the slice.  When termination occurs for the latter
reason, the final point will necessarily be inside the slice, and the
trajectory will therefore be accepted.

\section{A Demonstration}\label{sec-demo}\vspace*{-10pt}

To illustrate the benefits stemming from the adaptive nature of slice
sampling, I show here how it can help avoid a disastrous scenario, in
which a seriously wrong answer is obtained without any obvious
indication that something is amiss.

The task is to sample from a distribution for ten real-valued
variables, $v$ and $x_1$ to $x_9$.  The marginal distribution of
$v$ is Gaussian with mean zero and standard deviation~3.  Conditional
on a given value of $v$, the other variables, $x_1$ to $x_9$, are
independent, with the conditional distribution for each being Gaussian
with mean zero and variance $e^v$.  The resulting shape resembles a
ten-dimensional funnel, with small values for $v$ at its narrow end,
and large values for $v$ at its wide end.  Such a distribution is
typical of priors for components of Bayesian hierarchical models ---
$x_1$ to $x_9$ might, for example, be random effects for nine
subjects, with $v$ being the log of the variance of these random
effects.  If the data happens to be largely uninformative, the problem
of sampling from the posterior will be similar to that of sampling
from the prior, so this test is relevant to actual Bayesian inference
problems.

It is of course possible to sample from this distribution directly, by
simply sampling for $v$, and then sampling for each of $x_1$ to $x_9$
given this value for $v$, thereby obtaining independent points from
exactly the correct distribution.  And in any case, we already know
the correct marginal distribution for $v$, which will be the main
focus below.  For this test, however, we will pretend that we don't
already know the answer, and then compare what we would conclude using
various Markov chain methods to what we know is actually correct.

\begin{figure}[p]

\begin{center} \em Multivariate Metropolis updates, standard deviation 1 
\end{center}
\vspace*{-35pt}
\hspace*{-20pt}\psfig{figure=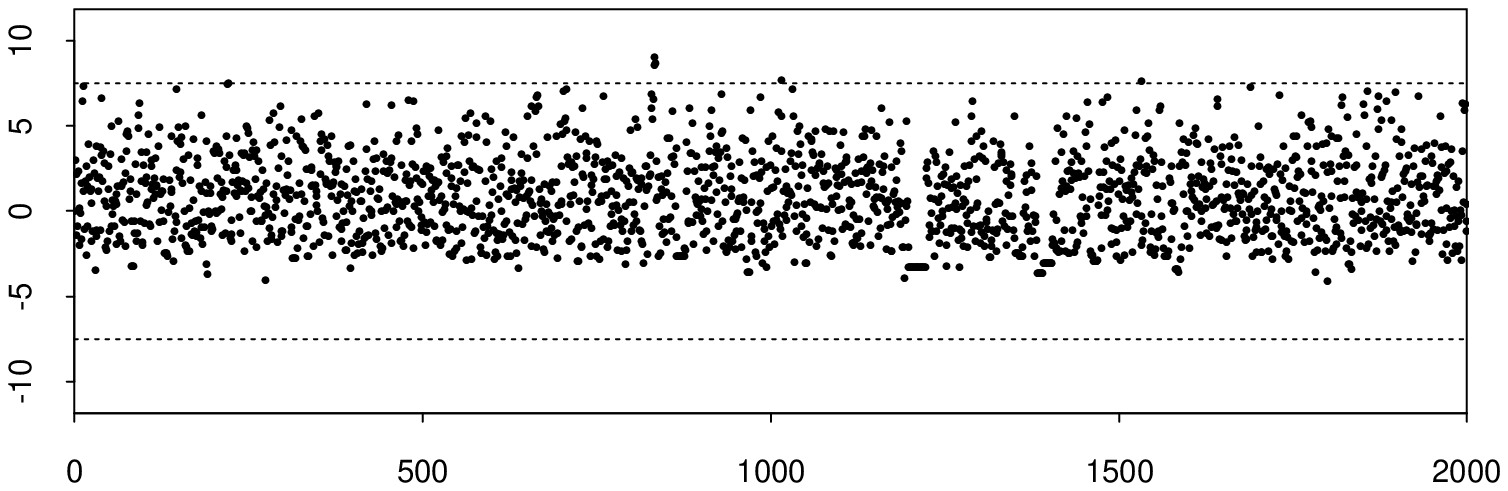}
\vspace*{-25pt}

\begin{center} \em Single-variable Metropolis updates, standard deviation 1 
\end{center}
\vspace*{-35pt}
\hspace*{-20pt}\psfig{figure=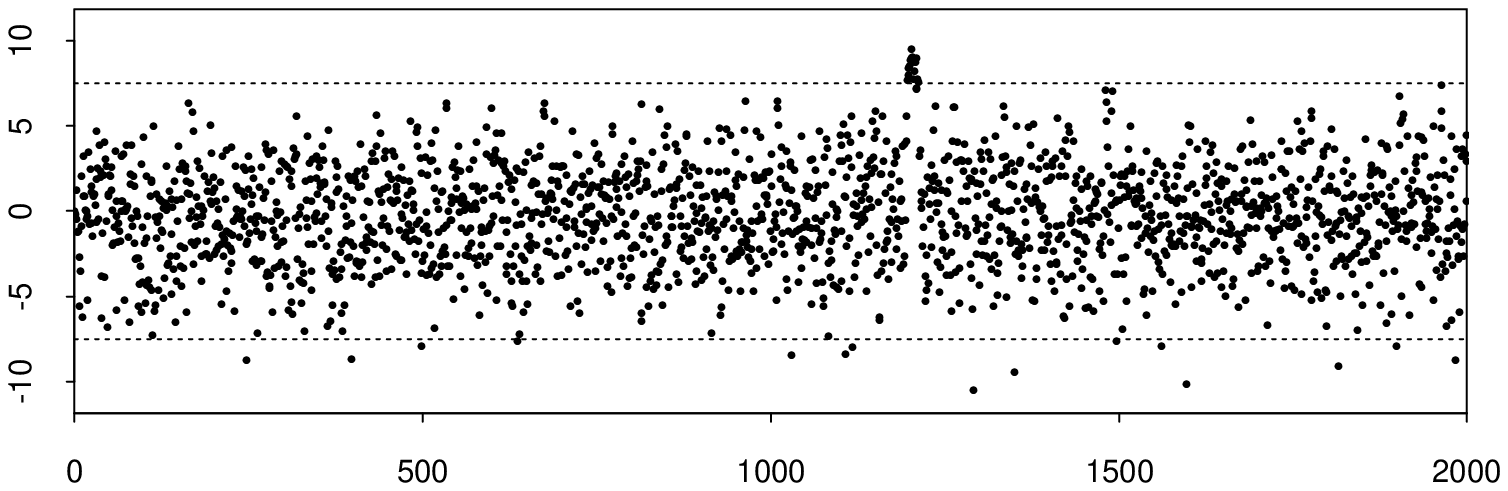}
\vspace*{-25pt}

\begin{center} \em Multivariate Metropolis updates, random standard deviation
\end{center}
\vspace*{-35pt}
\hspace*{-20pt}\psfig{figure=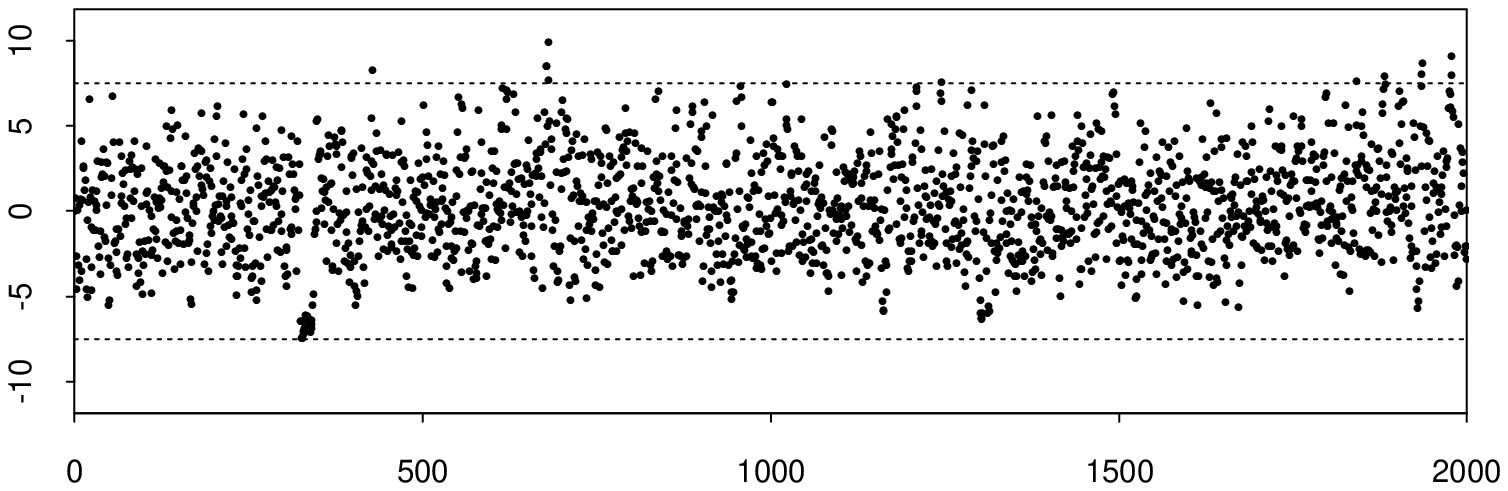}
\vspace*{-40pt}

\caption[]{Sampling from the funnel distribution with Metropolis 
           methods.}\label{fig-demo-met}

\end{figure}

Figure~\ref{fig-demo-met} shows the results of trying to sample from
this distribution using Metropolis methods.  The upper plot shows
2000 iterations of a run in which each iteration consists of 10000
multivariate Metropolis updates (ie, 20 million Metropolis updates
altogether).  The proposal distribution used was a spherical Gaussian
centred on the current state, with standard deviation of one for each
of the ten variables.  The initial state had $v=0$ and all $x_i=1$,
which is a typical magnitude for the $x_i$ given that $v=0$.  The
points plotted are the value of $v$ at each iteration.  Dotted lines
are shown at $v=\pm7.5$.

The results of this run are grossly incorrect.  We know that the
marginal distribution for $v$ is Gaussian with mean zero and standard
deviation~3.  One would expect that out of 2000 points from this
distribution, on average 95.6 (4.8\%) should be less than -5, but none
of the points sampled by the multivariate Metropolis method are in
this region.  Moreover, there is little in the plot to indicate that
anything is wrong.  In an actual application, the results of a run
such as this could easily be accepted as being correct, with serious
consequences.

The source of the problem is the low probability of accepting a
proposal when in a state where $v$ is small.  When $v$ is $-4$, for
example, the standard deviation of the $x_i$ conditional on this value
for $v$ is 0.135.  The chances that a multivariate Metropolis proposal
in which each $x_i$ has standard deviation one will produce values for
all the $x_i$ that are within this range of zero is about $0.135^9
\approx 1.5\times10^{-8}$.  The proposal will include a change to $v$
as well as the $x_i$, so this calculation does not give the exact
acceptance probability, but it does indicate that when $v$ is small,
the acceptance probability can become very small, and the chain will
remain in the same state for a very long time.  Since the Markov chain
leaves the correct distribution invariant, it follows that the chain
will only very rarely move from a large value of $v$ (which happens to
be where this run was started) to a small value for $v$ --- indeed,
this never occurred in the actual run.

Once one suspects a problem of this sort, signs of it can be seen in
the plot.  In particular, starting at iteration 1198, the value of $v$
stays at around $-3.3$ for 25 iterations (ie, for 250,000 Metropolis
updates).  However, there are no obvious occurrences of this sort in
the first 1000 iterations, so the problem would not be apparent even
to a suspicious user if only half as many iterations had been done.
Running several chains from different starting states might have
revealed the problem, but when sampling from more complex
distributions, it is difficult to be sure that an appropriate variety
of starting states has been tried.

The middle plot in Figure~\ref{fig-demo-met} shows the results of
sampling from the funnel distribution using single-variable Metropolis
updates, applied to each variable in sequence.  The proposal
distribution was a Gaussian centred on the current value, with
standard deviation one.  Each iteration for this run consisted of 1300
updates for each variable in turn, which take approximately as long as
10000 multivariate Metropolis updates (with the program and machine
used).  As before, the plot shows the value of $v$ after each of 2000
such iterations.

The results using single-variable Metropolis updates are not as
grossly wrong as those obtained using multivariate Metropolis updates.
Small values for $v$ are obtained in the expected proportion.  The
previous problem of very low acceptance rates when $v$ is small is
avoided because even when the standard deviation for one of the $x_i$
given $v$ is much smaller than the proposal standard deviation,
proposals to change a single $x_i$ are still accepted occasionally
(eg, when $v=-9$, the standard deviation of the $x_i$ is $0.011$, and
about one proposal in 100 is accepted).

However, \textit{large} values for $v$ are sampled poorly in this run.
About 0.6\% of the values should be greater than 7.5 (which is marked
by a dotted line), but no such values are seen in the first half of
the run (1000 iterations, 1.3 million updates for each variable).
Around iteration 1200, the chain moves to large values of $v$ and
stays there for 17 iterations (22100 updates for each variable).  This
number of points above 7.5 is not too far from the expected number in
2000 iterations, which is 12.4, so in this sense the run produced
approximately the right answer.  However, it is clear that this was
largely a matter of luck.  Movement to large values of $v$ is rare,
because once such a value for $v$ is reached, the chain is likely to
stay at a large value for $v$ for a long time.  In this case, the
problem is not a high rejection rate, but rather slow exploration of
the space in small steps.  For example, the standard deviation of the
$x_i$ when $v$ is 7.5 is 42.5.  Exploring a range of plus or minus
twice this by a random walk with steps of size around one takes about
$(4\times42.5)^2=28900$ updates of each variable.  While exploring
this range, substantial amounts of time will be spent with values for
the $x_i$ that are not compatible with smaller values of $v$.  (This
problem is not as severe in the previous run, because the multivariate
proposals take larger steps, since they change all variables at once.)

We might try to avoid the problems sampling for both large and small
values of $v$ by picking the proposal standard deviation at random,
from a wide range.  The lower plot in Figure~\ref{fig-demo-met} shows
the results when using multivariate Metropolis proposals in which the
log base 10 of the proposal standard deviation is chosen uniformly
from the interval $(-3,3)$.  Large values for $v$ are sampled fairly
well, but small values for $v$ are still a problem, though the results
are not as bad as for multivariate Metropolis with the proposal
standard deviation fixed at one.  Increasing the range of proposal
standard deviations to even more than six orders of magnitude might
fix the problem, but at an even greater cost in wasted computation
when the random choice is inappropriate.

\begin{figure}[t]

\begin{center} \em Single-variable slice sampling, initial width of 1
\end{center}
\vspace*{-35pt}
\hspace*{-20pt}\psfig{figure=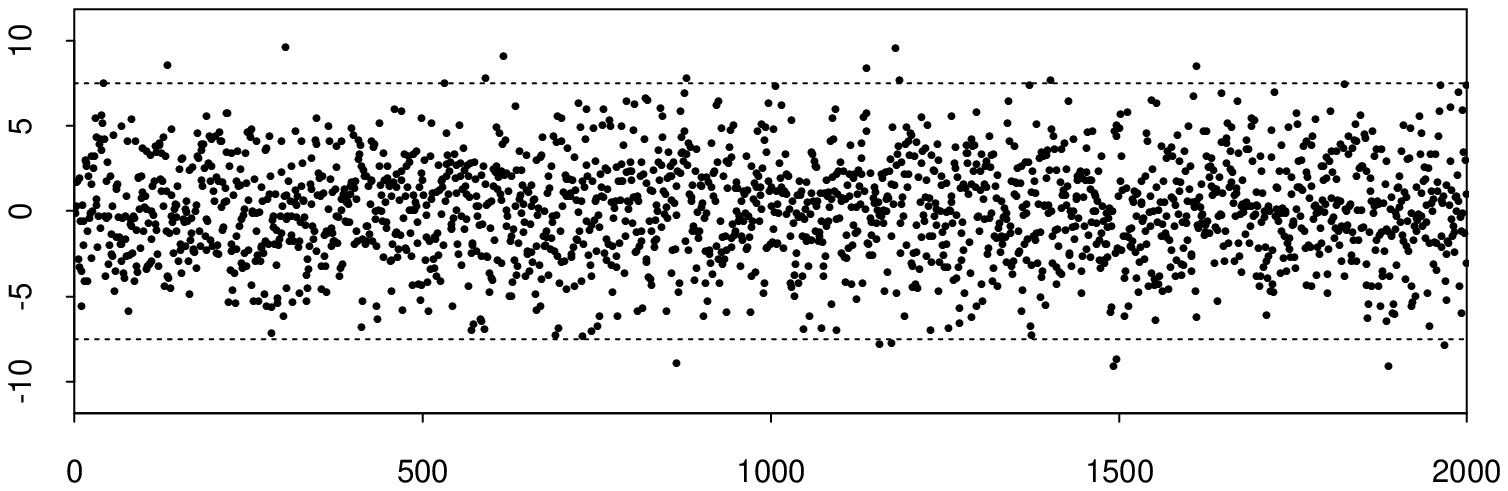}
\vspace*{-25pt}

\begin{center} \em Multivariate slice sampling, initial width of 1
\end{center}
\vspace*{-35pt}
\hspace*{-20pt}\psfig{figure=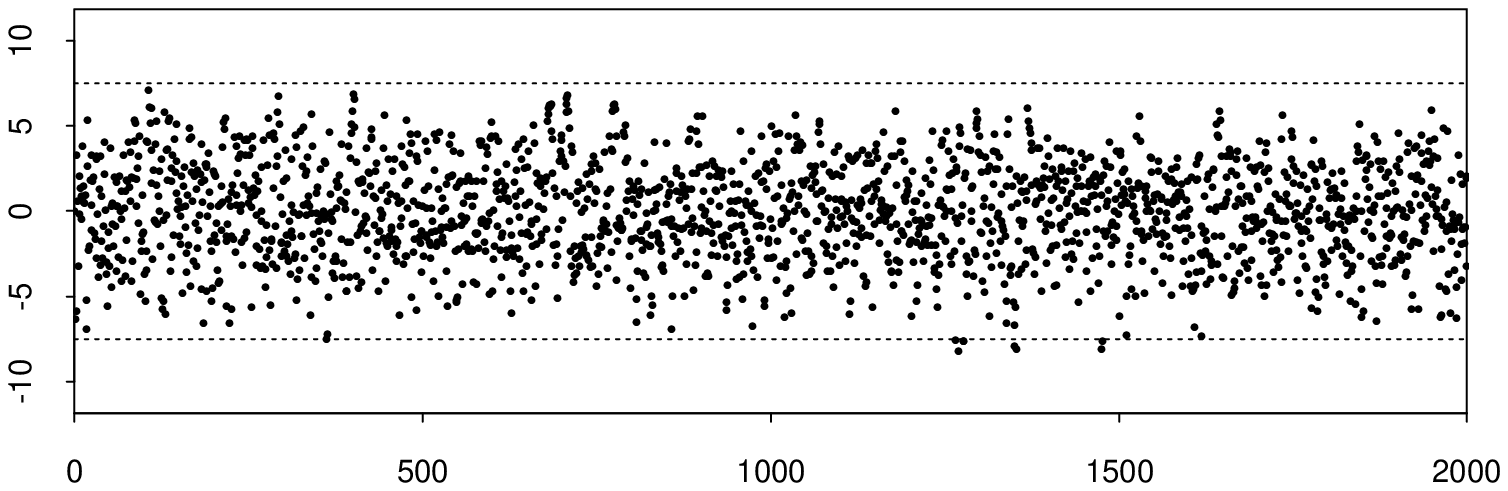}
\vspace*{-40pt}

\caption[]{Sampling from the funnel distribution with slice sampling
           methods.}\label{fig-demo-slc}

\end{figure}

Figure~\ref{fig-demo-slc} shows the results of trying to sample from
the funnel distribution using slice sampling methods.  In the upper
plot, single-variable slice sampling was used, with an initial
interval of size one, expanded by the stepping-out procedure
(Figure~\ref{fig-step}) until both ends are outside the slice, and
then sampled from with the shrinkage procedure
(Figure~\ref{fig-samp}).  Each of the 2000 iterations done consisted
of 120 such updates for each variable in turn, which takes
approximately the same amount of time as the Metropolis methods of
Figure~\ref{fig-demo-met}.  The average number of evaluations of $f$
for these slice sampling updates was $12.7$, but a few updates required
more than a hundred evaluations.

The results with single-variable slice sampling are quite good.  Small
values of $v$ are perhaps sampled slightly less well than with
single-variable Metropolis updates (Figure~\ref{fig-demo-met}, middle
plot), but the difference is not large.  Large values of $v$ are
sampled better than with any of the Metropolis methods.

The lower plot in Figure~\ref{fig-demo-slc} shows the results when
using the multivariate slice sampling procedure of
Figure~\ref{fig-simph}, with an initial hyperrectangle with sides of
length one.  Sampling is good out to values of $v$ of about $\pm7.5$,
but the extreme tails are not sampled well.  Recall that this
procedure does not expand the initial hyperrectangle, which explains
the poor sampling for large values of $v$.  The problem with small
values of $v$ is probably due to shrinkage being done for all
dimensions of the hyperrectangle, with the result that changes to $v$
are very small when $v$ is small (since changes to the $x_i$ must then
be small).  As mentioned briefly in Section~\ref{sec-hyper}, ways to
improve the method in both respects could be explored.

\section{Discussion}\label{sec-disc}\vspace*{-10pt}

As seen in this paper, the idea of slice sampling can be used to
produce many Markov chain sampling schemes.  In
Figure~\ref{fig-table}, I attempt to summarize the characteristics of
these schemes, and of some competing approaches for sampling from
general distributions on continuous state spaces.

\begin{figure}[b]
\begin{tabular}{lcccc}\hline\\[-9pt]
           & \em Derivatives
           & \em How critical 
           & \em Retrospective
           & \em Can suppress \\
           & \em needed?
           & \em is tuning? 
           & \em tuning allowed?
           & \em $\!\!$random walks?$\!\!$ \\[3pt]
\hline\\[-8pt]
\em Single-variable methods\hspace*{-3in} \\[3pt]
\hline\\[-5pt]
ARS/ARMS       & No (but helpful) & $\!\!$Low/Medium$\!\!$ & If log concave & No
\\[10pt]
Single-variable    & No               & Medium  & No             & No  \\
Metropolis         &                  &         &                &     \\[10pt]
Single-variable    & No               & Low     & If unimodal    & No  \\
slice sampling     &                  &         &                &     \\[10pt]
Overrelaxed        & No (but helpful) & Low     & $\!\!$If unimodal and$\!\!$
                   & Yes\\
slice sampling     &                  &       & $\!$endpoints exact$\!$&\\[10pt]
\hline\\[-8pt]
\em Multivariate methods\hspace*{-3in} \\[3pt]
\hline\\[-5pt]
Multivariate       & No               & Medium-High  & No             & No  \\
Metropolis         &                  &         &                &     \\[10pt]
Dynamical methods\hspace*{-10pt}
                   & Yes              & High    & No             & Yes \\[10pt]
Slice sampling with& No               & Low--Medium & No         & No  \\
 hyperrectangles   &                  &         &                &     \\[10pt]
Slice sampling with& Possibly         & Low--Medium & No          & No  \\
Gaussian crumbs    & helpful          &         &                &     \\[10pt]
Reflective         & Yes              & $\!\!\!\!\!$Medium--High$\!\!\!\!\!$  
                   & No          & Yes \\
slice sampling     &                  &         &                &     \\[3pt]
\hline
\end{tabular}

\vspace*{2pt}

\caption[]{Characteristics of some general-purpose Markov chain sampling 
           methods.}\label{fig-table}
\end{figure}

The table separates single-variable methods that update each variable
in turn from multivariate methods that update all variables at once.
Single-variable methods may be preferred when the coordinate system
used is such that one expects many of the variables to be almost
independent, allowing these variables to change by a large amount even
when the other variables are fixed.  Also, for some models,
recomputing the probability density after a change to one variable may
be much faster than recomputing it after a change to all variables.
On the other hand, if there are strong dependencies between variables,
using single-variable updates may lead to slow convergence, or even a
lack of ergodicity --- though for high-dimensional problems with
strong dependencies, simple-minded multivariate methods will also be
quite slow.

The first column in the table indicates whether the method requires
that derivatives of the (unnormalized) probability density be
computable.  Derivatives are needed by dynamical methods and
reflective slice sampling, which limits their applicability.  Adaptive
rejection sampling (Gilks and Wild 1992; Gilks 1992) and overrelaxed
slice sampling can take advantage of derivatives, but can operate
without such information with only a moderate loss of efficiency ---
eg, when no derivatives are available, overrelaxed slice sampling can
use bisection rather than Newton iteration to find the endpoints of
the slice.

The second and third columns indicate how critical it is that tuning
parameters be set to good values, and whether or under what conditions
`retrospective tuning' is allowed --- that is, whether parameters of
the method can be set based on information from past iterations.
Adaptive rejection sampling (ARS) for log concave distributions is
very good in these respects --- a parameter is needed for the size of
the first step taken in search of a point on the other size of the
mode, but subsequent steps can be made larger (eg, by doubling), so
the effect of a poor initial step is not too serious; furthermore, it
is allowable to set this size parameter based on the stepsize that was
found to be necessary in previous iterations.  Parameter tuning is
more of a problem when ARMS (Gilks, Best, and Tan 1995) is used for
distributions not known to be log concave --- a poor choice of
parameters may have worse effects, and retrospective tuning is not
allowed (Gilks, Neal, Best, and Tan 1997).  Tuning is also a problem
for single-variable and multivariate Metropolis methods --- proposing
changes that are too small leads to an inefficient random walk, while
proposing changes that are too large leads to frequent rejections.
Using too small a stepsize with a dynamical method is not quite as
bad, since movement is not in a random walk, but too large a stepsize
is disastrous, since the dynamical simulation becomes unstable, and
very few changes are accepted.  For Metropolis and dynamical methods,
the stepsize parameter must not be set retrospectively.

Single-variable slice sampling and overrelaxed slice sampling offer
advantages over other methods in these respects.  Whereas ARS/ARMS
allows retrospective tuning only for log concave distributions, this
is allowed for these slice sampling methods when they are applied to
any unimodal distribution (provided the interval is expanded to the
whole slice, and endpoints for overrelaxation are computed exactly).
Furthermore, the tuning is less critical for slice sampling than for
the other methods (apart from ARS), as discussed further below.  For
reflective slice sampling, however, tuning is at least moderately
critical, though perhaps less so than for dynamical methods, and
retrospective tuning is not allowed.  Tuning for multivariate slice
sampling using hyperrectangles is less critical than for multivariate
Metropolis methods, but as was seen in the demonstration of
Section~\ref{sec-demo}, tuning can be more critical for multivariate
slice sampling than for single-variable slice sampling.

The final column indicates whether the method can potentially suppress
random walk behaviour.  This is important when sampling from a
distribution with high dependencies between variables, as in such a
situation, exploration of the distribution may have to proceed in
small steps, and the difference in efficiency between diffusive and
systematic exploration of the distribution can be very large (as is
typical, for example, with neural network models (Neal 1996)).

Another way of exploring the differences between these methods is to
see how well they work in various circumstances.  The most favourable
situation is when our prior knowledge lets us choose good tuning
parameters for all the methods (eg, the width of a Metropolis proposal
distribution or of the initial interval for slice sampling).  A
Metropolis algorithm with a simple proposal distribution will then
move about the distribution fairly efficiently (although in a random
walk), and will have low overhead, since it requires evaluation of
$f(x)$ at only a single new point in each iteration.  Single-variable
slice sampling will be comparably efficient, however, provided we
stick with the interval chosen initially (ie, we set $m=1$ in the
stepping out procedure of Figure~\ref{fig-step}).  There will then be
no need to evaluate $f(x)$ at the boundaries of the interval, and if
the first point chosen from this interval is within the slice, only a
single evaluation of $f(x)$ will be done.  If this point is outside
the slice, further evaluations will be required, but this inefficiency
corresponds to the possibility of rejection with the Metropolis
algorithm.  This situation is similar for multivariate slice sampling
with an initial hyperrectangle that is not expanded.  Metropolis and
slice sampling methods should therefore perform quite similarly.
However, slice sampling will work better if it turns out that we
mistakenly chose too large a width for the Metropolis proposal
distribution and the initial slice sampling interval.  This error will
lead to a high rejection rate for the Metropolis algorithm, but the
sampling procedures of Figures~\ref{fig-samp} and~\ref{fig-simph} use
the rejected points to shrink the interval, which is much more
efficient when the initial interval was too large.

As seen in the demonstration of Section~\ref{sec-demo}, the advantage
of slice sampling over Metropolis methods can be quite dramatic if we
don't know enough to choose a good tuning parameter, or if no single
value of the tuning parameter is appropriate for the entire
distribution.

Another possibility is that we know that the conditional distributions
are log concave, but we do not know how wide they are.  Adaptive
Rejection Sampling (ARS) then works very well, because its width
parameter can be retrospectively tuned, based on previous iterations.
Single-variable slice sampling will also work well, since in this
situation it can also be tuned retrospectively (provided no limit is
set on the size of the interval).  However, ARS does true Gibbs
sampling, whereas the slice sampling updates do not produce points
that are independent of the previous point.  Such dependency is
probably a disadvantage (unless deliberately directed to useful ends,
as in overrelaxation), so ARS is probably better than single-variable
slice sampling in this context.

Suppose, however, that we know only that the conditional distributions
are unimodal, but not necessarily log concave.  We would then need to
use ARMS rather than ARS, and would not be able to tune it
retrospectively, whereas we can still use single-variable slice
sampling with retrospective tuning.  This will likely not be as good
as true Gibbs sampling, however, which we should prefer if the
conditional distribution happens to be one that can be efficiently
sampled from.  In particular, if slice sampling is used to sample from
a heavy-tailed distribution, it may move only infrequently between the
tails and the central region, since this transition can occur only
when we move to a point under the curve of $f(x)$ that is as low as
the region under the tails, but whose horizontal position is in the
central region.  However, there appears to be no general purpose
scheme that avoids problems in this situation.

Finally, consider a situation where we do not know that the
conditional distributions are unimodal, and have only a rough idea of
an appropriate width parameter for a proposal distribution or initial
slice sampling interval.  Single-variable slice sampling copes fairly
well with this uncertainty.  If the initial interval is too small it
can be expanded as needed, either by stepping out or by doubling
(which is better will depend on whether the faster expansion of
doubling is worth the extra overhead from the acceptance test of
Figure~\ref{fig-test}).  If instead the initial interval is too big,
it will be shrunk efficiently by the procedure of
Figure~\ref{fig-samp}.  We might try to achieve similar robustness
with the Metropolis algorithm by doing several updates for each
variable, using proposal distributions with a range of widths.  For
example, if $w$ is our best guess at an appropriate width, we might do
updates with widths of $w/4$, $w/2$, $w$, $2w$, and $4w$.  This may
ensure that an appropriate proposal distribution is used some of the
time, but it is unattractive for two reasons.  First, the limits of
the range (eg, from $w/4$ to $4w$) must be set \textit{a priori}.
Second, for this approach to be valid, we must continue through the
original sequence of widths even after it is clear that we have gone
past the appropriate one.  These problems are not present with slice
sampling.

In any of these situations, we might prefer to use a method that can
suppress random walks.  Dynamical methods such as Hybrid Monte Carlo
(Duane, \textit{et al} 1987) do this well for a wide range of
distributions; reflective slice sampling may also work for a wide
range of distributions, but preliminary indications are that is less
efficient than Hybrid Monte Carlo, when both are tuned optimally.
Overrelaxation is sometimes beneficial, but not always (whether it is
or not depends on the types of correlation present).  For problems
where overrelaxation is helpful, overrelaxed slice sampling may often
be the best approach to suppressing random walks.  If the conditional
distributions are unimodal, it offers the possibility of retrospective
tuning.  It does not require computation of derivatives.  For some
models, the fact that overrelaxation updates one variable at a time
will permit computational saving, in comparison with the simultaneous
updates for dynamical and reflective methods.

Multivariate slice sampling using hyperrectangles does not appear to
offer much, if any, advantage over single-variable slice sampling,
except for the uncommon situation were it is known that the coordinate
system used is especially bad (and hence updating variables singly
will be particularly inefficient).  However, the more general
framework for multivariate slice sampling based on `crumbs' that was
outlined in Section~\ref{sec-multiadapt} offers the possibility of
adapting not just to the scales of the variables, but also to the
dependencies between them.  The benefits of such methods can only be
determined after further research, but huge increases in efficiency
would seem conceivable, if one is to judge from the analogous
comparison of minimization by simple steepest descent versus more
sophisticated quasi-Newton or conjugate gradient methods.

The practical utility of the slice sampling methods described here
will ultimately be determined by experience on a variety of
applications.  Some such applications will involve tailor-made
sampling schemes for particular models --- for instance, Frey (1997)
used slice sampling successfully to sample for latent variables in a
neural network.  Slice sampling is also particularly suitable for use
in automatically generated samplers, and is now used in some
situations by the WinBUGS system (Lunn, {\em et al} 2000).  Readers
can try out slice sampling methods for themselves, on a variety of
Bayesian models, using the ``software for flexible Bayesian modeling''
that is available from my web page.  This software (version of
2000-08-21) implements most of the methods discussed in this paper.

\section*{Acknowledgements}\vspace*{-10pt}

I thank Brendan Frey, Gareth Roberts, Jeffrey Rosenthal, and David
MacKay for helpful discussions.  This research was supported by the
Natural Sciences and Engineering Research Council of Canada and by
the Institute for Robotics and Intelligent Systems.

\section*{References}\vspace*{-10pt}

\leftmargini 0.2in
\labelsep 0in

\begin{description}
\itemsep 2pt

\item
  Adler, S.~L.\ (1981) ``Over-relaxation method for the Monte Carlo evaluation
  of the partition function for multiquadratic actions'', {\em Physical
  Review D}, vol.~23, pp.~2901-2904.

\item
  Barone, P.\ and Frigessi, A.\ (1990) ``Improving stochastic relaxation
  for Gaussian random fields'', {\em Probability in the Engineering and
  Informational Sciences}, vol.~4, pp.~369-389.

\item
  Besag, J. and Green, P. J. (1993) ``Spatial statistics and Bayesian
  computation'' (with discussion), {\em Journal of the Royal
  Statistical Society B}, vol.~55, pp.~25-37 (discussion, pp.~53-102).

\item
  Damien, P., Wakefield, J., and Walker, S.\ (in press) ``Gibbs sampling
  for Bayesian nonconjugate and hierarchical models using auxiliary
  variables'', to appear in {\em The Journal of the Royal Statistical
  Society B}.

\item
  Diaconis, P., Holmes, S., and Neal, R.~M.\ (in press) ``Analysis of a 
  non-reversible Markov chain sampler'', to appear in {\em The
  Annals of Applied Probability}.

\item
  Duane, S., Kennedy, A.~D., Pendleton, B.~J., and Roweth, D. (1987)
  ``Hybrid Monte Carlo'', {\em Physics Letters B}, vol.~195, pp.~216-222.

\item
  Edwards, R.~G.\ and Sokal, A.~D.\ (1988) ``Generalization of the 
  Fortuin-Kasteleyn-Swendsen-Wang representation and Monte Carlo
  algorithm'', {\em Physical Review D}, vol.~38, pp.~2009-2012.

\item
  Frey, B.~J.\ (1997) ``Continuous sigmoidal belief networks trained
  using slice sampling'', in M.~C.~Mozer, M.~I.~Jordan, and T.~Petsche
  (editors) {\em Advances in Neural Information Processing Systems 9},
  MIT Press.

\item
  Gelfand, A.~E.\ and Smith, A.~F.~M. (1990) ``Sampling-based
  approaches to calculating marginal densities'', {\em Journal
  of the American Statistical Association}, vol.~85, pp.~398-409.


\item
  Gilks, W.~R.\ (1992) ``Derivative-free adaptive rejection sampling
  for Gibbs sampling'', in J.~M.\ Bernardo, J.~O.\ Berger, A.~P.\ Dawid,
  and A.~F.~M.\ Smith (editors), {\em Bayesian Statistics 4}, pp.~641-649,
  Oxford University Press.

\item
  Gilks, W.~R., Best, N.~G., and Tan, K.~K.~C.\ (1995) ``Adaptive
  rejection Metropolis sampling within Gibbs sampling'', {\em Applied
  Statistics}, vol.~44, pp.~455-472.

\item
  Gilks, W.~R., Neal, R.~M., Best, N.~G., and Tan, K.~K.~C.\ (1997)
  ``Corrigendum: Adaptive rejection Metropolis sampling'', 
  {\em Applied Statistics}, vol.~46, pp.~541-542.

\item
  Gilks, W.~R.\ and Wild, P.\ (1992) ``Adaptive rejection sampling
  for Gibbs sampling'', \textit{Applied Statistics}, vol.~41, pp.~337-348.

\item
  Green, P.~J.\ and Han, X.\ (1992) ``Metropolis methods, Gaussian
  proposals and antithetic variables'', in P. Barone, {\em et al.\/}\
  (editors) {\em Stochastic Models, Statistical Methods, and Algorithms 
  in Image Analysis}, Lecture Notes in Statistics, Berlin: Springer-Verlag.

\item
  Green, P.~J.\ and Mira, A.\ (1999) ``Delayed rejection in reversible
  jump Metropolis-Hastings'', Mathematics Research Report S-01-99, 
  University of Bristol.

\item
  Hastings, W.~K.\ (1970) ``Monte Carlo sampling methods using Markov chains 
  and their applications'', {\em Biometrika}, vol.~57, pp.~97-109.

\item
  Higdon, D.~M.\ (1996) ``Auxiliary variable methods for Markov chain
  Monte Carlo with applications'', ISDS Discussion Paper 96-17, 25 pages.

\item
  Horowitz, A.~M. (1991) ``A generalized guided Monte Carlo algorithm'',
  {\em Physics Letters B}, vol.~268, pp.~247-252.

\item
  Lunn, D.~J., Thomas, A., Best, N., and Spiegelhalter, D.\ (2000)
  ``WinBUGS -- a Bayesian modelling framework: concepts, structure,
  and extensibility'', {\em Statistics and Computing}, vol.~10, pp.~321-333.

\item
  Metropolis, N., Rosenbluth, A.~W., Rosenbluth, M.~N., Teller, A.~H., 
  and Teller, E.\ (1953) ``Equation of state calculations by fast computing 
  machines'', {\em Journal of Chemical Physics}, vol.~21, pp.~1087-1092.

\item
  Mira, A.\ (1998) {\em Ordering, Splicing, and Splitting Monte Carlo
  Markov Chains}, Ph.D.\ Thesis, School of Statistics, University of 
  Minnesota.

\item
  Mira, A.\ and Tierney, L.\ (in press) ``On the use of auxiliary variables
  in Markov chain Monte Carlo sampling'', to appear in {\em The 
  Scandinavian Journal of Statistics}.

\item
  Neal, R.~M.\ (1993) {\em Probabilistic Inference Using Markov Chain
  Monte Carlo Methods}, Technical Report CRG-TR-93-1, Dept.\
  of Computer Science, University of Toronto, 140 pages.  Obtainable 
  from \texttt{http://www.cs.utoronto.ca/$\sim$radford/}.

\item
  Neal, R.~M.\ (1994) ``An improved acceptance procedure for the 
  hybrid Monte Carlo algorithm'', {\em Journal of Computational Physics},
  vol.~111, pp.~194-203.

\item
  Neal, R.~M.\ (1998) ``Suppressing random walks in Markov chain Monte Carlo 
  using ordered overrelaxation'', in M.~I.~Jordan (editor), 
  {\em Learning in Graphical Models}, pp.~205-228, Dordrecht: Kluwer 
  Academic Publishers.

\item
  Neal, R. M. (1996) \textit{Bayesian Learning for Neural Networks},
  Lecture Notes in Statistics No.~118, New York: Springer-Verlag.

\item
  Roberts, G.~O.\ and Rosenthal, J.~S.\ (1999) ``Convergence of 
  slice sampler Markov chains'', {\em Journal of the Royal Statistical
  Society B}, vol.~61, pp.~643-660.

\item 
  Swendsen, R.~H.\ and Wang, J.-S.\ (1987) ``Nonuniversal critical
  dynamics in Monte Carlo simulations'', {\em Physical Review Letters},
  vol.~58, pp.~86-88.

\item
  Tierney, L.\ and Mira, A.\ (1999) ``Some adaptive Monte Carlo methods 
  for Bayesian inference'', {\em Statistics in Medicine}, vol.~18, 
  pp.~2507-2515.

\item
  Thomas, A., Spiegelhalter, D.~J., and Gilks, W.~R.\ (1992) ``BUGS: A
  program to perform Bayesian inference using Gibbs sampling'', in J.~M.\
  Bernardo, J.~O.\ Berger, A.~P.\ Dawid, and A.~F.~M.\ Smith (editors), 
  {\em Bayesian Statistics 4}, pp.~837-842, Oxford University Press.

\end{description}
\end{document}